\documentclass{article}
\pdfoutput=1
\usepackage{graphicx}
\usepackage{rotate}
\usepackage{rotating}
\usepackage{algorithm}
\usepackage{algorithmic}
\usepackage{amsmath}
\usepackage{amsfonts}

\title{
The Invisible Hand Heuristic for Origin-Destination Integer
Multicommodity Network Flows}
      \author{Richard S. Barr\thanks{Department
      of Engineering Management, Information, and Systems,
      Lyle School of Engineering,
      Southern Methodist University, Dallas, TX 75275. Email:
      {\tt barr@smu.edu} (Corresponding author)} \and Thomas Mc{L}oud\thanks{Email: {\tt
      twmcloud@hotmail.com}.}
      }
      \date{\today }

\begin{document}
\maketitle

\begin{abstract}
Origin-destination integer multicommodity flow problems differ
from classic multicommodity models in that each commodity has one
source and one sink, and each commodity must be routed along a
single path. A new invisible-hand heuristic that mimics economic
markets' behavior is presented and tested on large-scale
telecommunications networks, with solution times two orders of magnitude faster than Cplex's LP relaxation, more dramatic MIP ratios, and small solution value differences.
\end{abstract}

\vspace{2in}
\noindent {\small {\emph{Keywords:} network optimization, multicommodity, integer programming, economics, heuristics, telecommunications, logistics}

\newpage
\section{Introduction}
The general problem for origin-destination integer muliticommodity
network flows (ODIMCF) consists of a network with limited capacity
on one or more arcs and several distinct, non-interchangeable commodities sharing this limited
network capacity to satisfy their respective demands and supplies.
Hence, all commodities have separate,  structurally identical networks with upper bounds on the sum of flows across corresponding arcs.
While this description is consistent with the classic minimum-cost multicommodity
network flow (MCF) problem
\cite{AHUJ93,ASSA78,BAZA90,KENN80,williamson2019nwflow},
ODIMCF has two  differentiating aspects:


\begin{enumerate}

\item Each commodity has a single source (supply node) and a
single sink (demand node).

\item The entire flow of each commodity must follow a single path
from its source (origin) to its sink (destination).

\end{enumerate}

\noindent This last requirement makes ODIMCF an integer
programming problem. As the number of commodities increases, the
size of an ODIMCF instance grows rapidly.  This combination of
large instances and integrality requirements increases the
difficulty of solving these problems.

This research is motivated by the presence of large instances of
ODIMCF models in practice, with hundreds of thousands of constraints and millions of binary variables,
which may require quick or repeated solution.
Even modest problem instances can challenge the effectiveness of current optimization
methodologies. The combination of these issues serves
as a strong motivator for the development of more efficient solution
techniques in terms of speed and solution quality.

This paper develops a new heuristic approach for the solution of
ODIMCF problems.  The algorithm has polynomial asymptotic bounds
for both space and time.  The minimal space requirement enables
the solution of large problem instances for which testing demonstrates 
 extremely small running times and near-optimal solutions.

\section{Applications and Literature Review}

Large instances of ODIMCF occur in communications, package distribution, computer,  transportation, 
supply-chain distribution, and traffic networks \cite{AHUJ91,BARN00,BAZA90}. In a transportation example, 
Huntley et al.\ \cite{HUNT95} describe a
problem from the railroad industry: the movement of loaded
grain cars that are grouped into blocks and  moved from their origins to their destinations. 
The grain trains connecting stations have load limits on total weight and
length and multiple blocks can share a train's capacity. 
The combination of a station and train arrival/departure times
forms network nodes and blocks traverse across arcs
representing track usage or waiting at a station. 
These nodes, arcs, and blocks form an
ODIMCF instance, with each block of freight cars treated as a separate commodity.

Traffic routing in multi-protocol label-switching (MPLS) and similar
network technologies, such as segment routing \cite{moreno2017segment}, is an instance of ODIMCF in the telecommunications industry
\cite{GIRI00,LITO99,resende2003grasp,SAIT00}.  
A label-switched path (LSP) is
established for groupings of traffic having the same origin and
destination in an  MPLS network. All traffic
assigned to an LSP will follow the same path across the network, yet
the LSPs share the limited network bandwidth, expressed as capacities on the arcs.  Girish et al.
\cite{GIRI00} provide a formulation for MPLS traffic routing
consistent with ODIMCF having LSPs serve as the commodities, along
with additional formulations for specializations of this problem.
The number of LSPs in even a small MPLS network can be large since  at
least one LSP may be required to connect each node to every other network node.
For example, in a small 30-node network, 800 or 900 LSPs
(commodities) are typical in practice.  

Table \ref{map} summarizes
the mapping of application components to ODIMCF elemenets for both MPLS and
grain-car movement applications.
Other ODIMCF applications similar to that of MPLS routing include: 
wavelength-division multiplexing in optical networks without bifurcated flow \cite{raghaven2011}, the Virtual Network Embedding Problem of mapping virtual communications networks with heterogeneous topologies onto a physical networks \cite{moura2018}, provisioning long-term private virtual circuits between customer endpoints on a
large backbone network \cite{resende2003grasp}, and satellite payload configuration  to optimize power usage while ensuring sufficient signal amplification for retransmission on the downlink \cite{KIEF2019}. 


\begin{table}[tbp]
  \begin{center}
  \caption{Mapping of Applications to ODIMCF} \footnotesize
\begin{tabular}{lll}

Application & MPLS & Grain-Car Movement \\ \hline

Commodities & LSPs & Blocks \\

Demand & LSP bandwidth & Block length \\

Nodes & Switches and routers & Train arrival
or departure at a station \\

Arcs & Network links & Remaining at a station or movement by train \\

Arc Capacities & Link bandwidth &  Maximum train length and
station capacity\\ \hline

\end{tabular} \label{map} \end{center} \end{table}


While these ODIMCF problems can be formulated as generic
integer programming models \cite{chen2010appliedIP,NEMH88,WOLS98}, realistic instances
are challenging to solve with current software and specialized approaches are warranted.   
Specialized exact algorithms have been developed by Barnhart et al.\ \cite{BARN00}, Park et al.\ \cite{PARK96}, and Moura et al.\ \cite{moura2018} that use
column-generation and branch-and-bound techniques to
solve small instances of ODIMCF.  These approaches use
price-directive decomposition to solve the linear programming
relaxations at the nodes in a branch-and-bound tree.  Cutting
planes are used at the nodes to improve the solutions found at
each node. 

But heuristic techniques have also been developed for these problems to
enable the solution of larger problem instances.
Early work by Huntley et al.\ \cite{HUNT95} utilizes simulated annealing
\cite{gendreau2010metaheur} to approximately solve an ODIMCF problem. Details of
the procedure are incomplete, but good results are claimed. Laguna and Glover
\cite{LAGU93} use Tabu search for the related bandwidth-packing problem. Resende and Ribeiro \cite{resende2003grasp} applied the
GRASP metaheuristic \cite{rese2016graspbook} to route private virtual circuits through a backbone telecommunications network.
Amiri et al.\ \cite{AMIR00,AMIR99}, Rolland et al.\ \cite{ROLL99}, and recently Fortz et al.
 \cite{fortz2017Lagrangian} present Lagrangian-relaxation-based heuristics for ODIMCF. And 
Brun et al. \cite{brun2017} develop an approximation heuristic inspired by game 
theory's Nash equilibrium

The following sections present a mathematical statement of the problem
and develop  a new heuristic  based on classic economic principles. Computational testing on large problem sets demonstrates the effectiveness of this approach.

\section{ Mathematical Formulation}

In formulating an ODIMCF problem, the network topology, arc
capacities, and commodity information are assumed to be
deterministic and given. Let $\mathbb{B} = \{0,1\}$ be the set of binary 
numbers, $\mathbb{R}$
be the set of real numbers, $\mathbb{R}_+$ be the set of positive
real numbers, and $\mathbb{Z}_{0+}$ be the set of non-negative
integers.

Define $K$, $N$, and $A$ to be the sets of
commodities, nodes, and directed arcs, respectively.  For directed arc $a \in 
A$,
let $c_a \in \mathbb{R}_{0+}$ be the non-negative cost per unit of
flow, $u_a \in \mathbb{R}_+$ be the capacity limit, and $i_a \in
N$ ($j_a \in N$) be the tail (head) of the arc. The
characteristics $c_a$, $u_a$, $i_a$, and $j_a$ are universally
associated with each $a \in A$. 

Within the network, a {\em route}, $P \subset A$, is a set of arcs with the 
following 
characterstics:


\begin{enumerate}

\item If $P \ne \emptyset$, then $P$ has an origin (destination) node $s \in N$ ($t \in N$) at which $P$ originates (terminates).

\item $\forall a \in P, \, j_a \ne t \Rightarrow \exists b \in P \text{ s.t. } j_a = i_b$.

\item $\forall a \in P, \, i_a \ne s \Rightarrow \exists b \in P \text{ s.t. } j_b = i_a$.

\item $P \ne \emptyset \Rightarrow \exists a \in P \text{ s.t. } i_a = s \, (j_a = t)$.

\item $\forall a \in P, \, \text{ there does not exist } b \in P \text{ s.t. } i_a = i_b \, (j_a = j_b)$.

\item If $P \ne \emptyset$, then the directed network formed by the directed arcs of $P$ and their heads and tails is a tree.

\end{enumerate}


\noindent 

Each commodity $k \in K$ has an origin $s_k \in N$, destination
$t_k \in N$, and required flow from $s_k$ to $t_k$ of $d_k \in
\mathbb{R}_{+}$. This demand for commodity $k$ is represented by
$d_k$ units of supply at $s_k$ and $d_k$ units of demand at $t_k$
indicated in the demand vector $\mathbf b^k$ with a 1 (-1) entry
corresponding to  $t_k$ ($s_k$) and 0 for all other nodes. The
characteristics $s_k$, $t_k$, $d_k$, and $\mathbf b^k$ are
universally associated with each $k \in K$.

Let $\mathbf X$ be a matrix of binary flow variables for all
commodities. If $\text{X}_{a,k}$ is 1 (0) then commodity $k$ uses
(does not use) arc $a \in A$. For node $n \in N$, let $E(n)$ be
the set of directed arcs emanating from $n$ and $T(n)$ be the set
of directed arcs terminating at $n$. Table \ref{opc} contains a summary
of the components of an ODIMCF problem.

\begin{table}[tbp]
	\begin{center}
		\caption{ODIMCF Problem Components} \footnotesize
		\begin{tabular}{lll}

			\multicolumn{1}{c}{Component} & \multicolumn{1}{c}{Type} & \multicolumn{1}{c}{Definition} \\ \hline
			
			$T(n) \subset A $ &  Constant & Set of arcs terminating at $n \in N$ \\
			$E(n) \subset A $ &  Constant & Set of arcs emanating from $n \in N$ \\
			$u_a \in \mathbb{R}_+$ & Constant & Capacity limit on total flow for $a \in A$ \\
			$c_a \in \mathbb{R}_{0+}$ &  Constant & Cost per unit of flow for $a \in A$ \\
			$i_a \in N$ & Constant & Node from which $a \in A$ emanates\\
			$j_a \in N$ &  Constant & Node at which $a \in A$ terminates\\
			$s_k \in N$ &  Constant & Origin or source node for $k \in K$\\
			$t_k \in N$ &  Constant & Destination or sink node for $k \in K$\\
			$d_k \in \mathbb{R}_+$& Constant & Demand (supply) for commodity $k \in K$ at $t_k$ ($s_k$)\\
			$b^k_n \in \mathbb{B} $ & Constant & $1 \Rightarrow n = t_k, -1 \Rightarrow n = s_k, 0 \, \mathrm{otherwise.} \, k \in K, \, n \in N$  \\
			$\text{X}_{a,k} \in \mathbb{B} $ & Variable & $1 (0) \Rightarrow$
			commodity $k \in K$ uses (does not use) arc $a \in A$ \\ \hline
		\end{tabular} \label{opc} \end{center} \end{table}

The node-arc formulation for ODIMCF is given by (\ref{odb1})--(\ref{odb4}).\footnote{For a path-based formulation, see \cite{BARN00,HUNT95}.}  The objective function, (\ref{odb1}), seeks to
minimize total routing cost for all commodities. The node-balance
equations, (\ref{odb2}), ensure that the flow of each commodity
satisfies the conservation of flow at the nodes and supply and
demand requirements. The limit on arc capacities is enforced
across all commodities in (\ref{odb3}). The integrality
requirements, (\ref{odb4}), require that the flows for each
commodity follow a single path through the network.

\begin{align}
 & \Big[ \text{ODIMCF} \Big] \nonumber\\
&\text{Minimize:}& \sum_{  k \in K} \sum_{a \in A} d_k c_a \text{X}_{a,k} &\label{odb1}\\
&\text{subject to:} & \sum_{a \in T(n)} \text{X}_{a,k} - \sum_{a \in E(n)} \text{X}_{a,k} &= b^k_n &&\forall n \in N ,\, \forall k \in K \label{odb2}\\
& &\sum_{ k \in K} d_k  \text{X}_{a,k}&\le u_a  &&\forall a \in A \label{odb3} \\
& & \text{X}_{a,k} &\in \mathbb{B}    &&\forall k \in K,\, \forall
a \in A \label{odb4}
\end{align}

\section{Invisible-Hand Heuristic for ODIMCF}

The solution of large instances of ODIMCF have proven to be challenging for standard
optimization techniques \cite{BARN00}.  With this as motivation, a
new heuristic is developed that quickly determines near-optimal
solutions for large-scale problems with many commodities.

Many successful metaheuristics are inspired by systems that
evolved naturally. Corne et al.\ \cite{CORN99} and Gendreau and Potvin \cite{gendreau2010metaheur} present many examples of
such approaches, which include genetic algorithms, immune-system
methods, ant-colony optimization, and particle swarm.  Garlick and
Barr \cite{GARL02} use ant-colony optimization for the routing and
wavelength assignment problem, which has many characteristics in
common with ODIMCF.

The new heuristic presented below is inspired by Adam Smith's
insights into market-based economic systems.  In 1776, Adam Smith
wrote the following \cite{SMITH}:

\begin{quote}
Every individual necessarily labours to render the annual revenue
of the society as great as he can.  He generally neither intends
to promote the public interest, nor knows how much he is promoting
it \dots\  He intends only his own gain, and he is in this, as in
many other cases, led by an {\em invisible hand} to promote an end
which was no part of his intention.  Nor is it always the worse
for society that it was no part of his intention.  By pursuing his
own interest he frequently promotes that of the society more
effectually than when he really intends to promote it.  I have
never known much good done by those who affected to trade for the
public good.
\end{quote}

Based on Smith's observation, the {\em invisible hand heuristic} (IHH) is designed to emulate and exploit the forces at work in a competitive marketplace. A specific application of this approach is developed for ODIMCF, wherein each commodity must choose a path over which to be routed. Just as the price mechanism is the control mechanism of a true market system \cite{bowden1989econ,mankiw2011econ,snyder2008econ}, IHH uses resource prices as its control mechanism, where the resources are the arc capacities. IHH's pricing mechanism consists of two components, the original arc costs and a heuristic {\em scarcity cost} unique to each arc. The {\em market cost} of an arc is the sum of scarcity cost and the original arc cost. The original arc costs are infinitely elastic, not varying with quantity of an arc’s capacity consumed by commodities. The scarcity cost function is designed to become increasingly inelastic as the quantity of an arc’s capacity is consumed---for each additional unit of capacity consumed the slope of the marginal cost function increases. The increasing resource price works with the commodity demand curves’ rationing function to help limit consumption of the scarce resources in the network---arc capacity. In this way, each arc is an independent monopolistic supplier of a unique resource and adjusts the price of it based solely on the current demand it sees for its capacity.

The commodities in the ODIMCF problem are the consumers of the resources with each trying to acquire a set of complimentary goods–--capacity on specific arcs---to form a path from its origin to its destination that minimizes the total market cost of the path. For a commodity, the arcs along a possible path from origin to destination are complimentary goods so that the price of capacity on one arc affects the commodity's demand for capacity on other arcs in the path. As the price for an arc's capacity goes down (up), the commodity's demand for capacity on complimentary arcs will go up (down). As each commodity has multiple paths to choose from in the network, arc capacity for arcs in alternate paths are substitute goods. As the price for capacity on an arc goes down (up) the commodity’s demand for capacity on substitute arcs will go down (up). As each commodity may have a different origin-destination pair with different possible paths, the set of complimentary and substitute goods will vary by commodity.

IHH does not attempt to determine the demand curve for each arc’s capacity. Nor does IHH have a central planner coordinating individual arc prices or allocating arc capacity to specific commodities. Instead, each commodity continuously attempts to minimize the market cost of its route as the costs change. This process of continuous reevaluation proceeds until an equilibrium is reached and all commodities are satisfied with their routes. The commodities never consider the effect of routing decisions on other commodities (the entire society); each commodity considers only its own self-interest. The only interaction between commodities and between arcs and commodities occurs through the price mechanism.

\subsection{Residual Capacity and Market Costs}

ODIMCF problems have hard limits on the availability of each resource---arc's capacity---and is a short-run problem where no additional capacity can be added. To satisfy the arc-capacity limits, the marginal market-cost curve is designed to reach an equilibrium point where the total capacity utilized by the commodities is less than the available supply. As ODIMCF is also trying to minimize total routing cost and not merely satisfy the capacity constraints, the marginal market cost curve must also reflect the original arc costs. 

The scarcity cost component of the marginal market cost is focused on achieving equilibrium and follows the law of diminishing returns with marginal cost rising as utilized capacity for an arc approaches the arc's capacity limit, $u_a$. The scarcity cost of an arc varies by commodity and is a function of the residual capacity of the arc and the $d_k$ for the commodity $k$. As defined in Table \ref{mktcosts}, let $r(a,k, \mathbf X) \rightarrow \mathbb{R}$ (\ref{res}) be the {\em residual 	capacity} available for commodity $k \in K$ on arc $a \in A$,  with no requirement that it be nonnegative. The scarcity cost function  $sc(a,k, \mathbf X)  \rightarrow
\mathbb{R}$  (\ref{sc}) reflects an increase in cost or price for commodity $k$ as residual capacity on arc $a$ approaches zero. The associated parameters $\beta$, $\mu$, and $\pi$ are positive real-valued scalars, each having the same value for all arcs and commodities. The values of these parameters are set a priori and are discussed further in Section \ref{bpm}. 

The marginal market cost of commodity $k$ on arc $a$  is determined by the function $mc(a,k, \mathbf X) \rightarrow \mathbb{R}_{0+}$
(\ref{mc}) and is a marginal cost in that it represents the cost of the last unit of flow if commodity $k$ were to use arc $a$. Function
$rc(P,k,A,\mathbf X) \rightarrow \mathbb{R}_{0+}$ (\ref{rtec}) defines the {\em marginal market cost} of route $P \subset A$.

Since $r(a,k, \mathbf X)$ changes based on the current route selection of the other commodities, $K \setminus \{k\}$, the
marginal market cost of an arc is not fixed and varies as the flows of other commodities change. As $sc(a,k, \mathbf X)$ is dependent
upon commodity $k$'s demand, $d_k$, market costs vary by commodity.
Figure \ref{mccurv} shows a graphical representation of $mc(a,k,
\mathbf X)$ as it relates to $r(a,k, \mathbf X)$ and $d_k$. In
the diagram, the current utilization corresponds to $u_a -
r(a,k, \mathbf X)$, the arc capacity utilized by other
commodities.  The market cost is determined as the intersection of
the resulting arc capacity utilization if commodity $k$ uses arc
$a$, $u_a - r(a,k, \mathbf X) + d_k$, and the marginal
market-cost curve.

\begin{table}
	\caption{Market and Scarcity Cost Definitions\label{mktcosts}}	
	\begin{align}
	r(a,k, \mathbf X) & = u_a - \sum_{g \in K \setminus \{k\}} d_g \text{X}_{a,g} 
	\label{res}\\
	sc(a,k, \mathbf X) &= \mu \left( max \left( 0,\frac{\beta + d_k -
		r(a,k, \mathbf X)}{\beta} \right) \right) ^{\pi} \label{sc}\\
	mc(a,k, \mathbf X) &= sc(a,k, \mathbf X) + c_a \label{mc}\\
	rc(P,k,A,\mathbf X)  &=   \sum_{ a \in P} mc(a,k, \mathbf X)
	\label{rtec}
	\end{align}
\end{table}

\begin{figure}[tbp]
	\begin{center}
		\includegraphics[angle = 270, scale = 0.5]{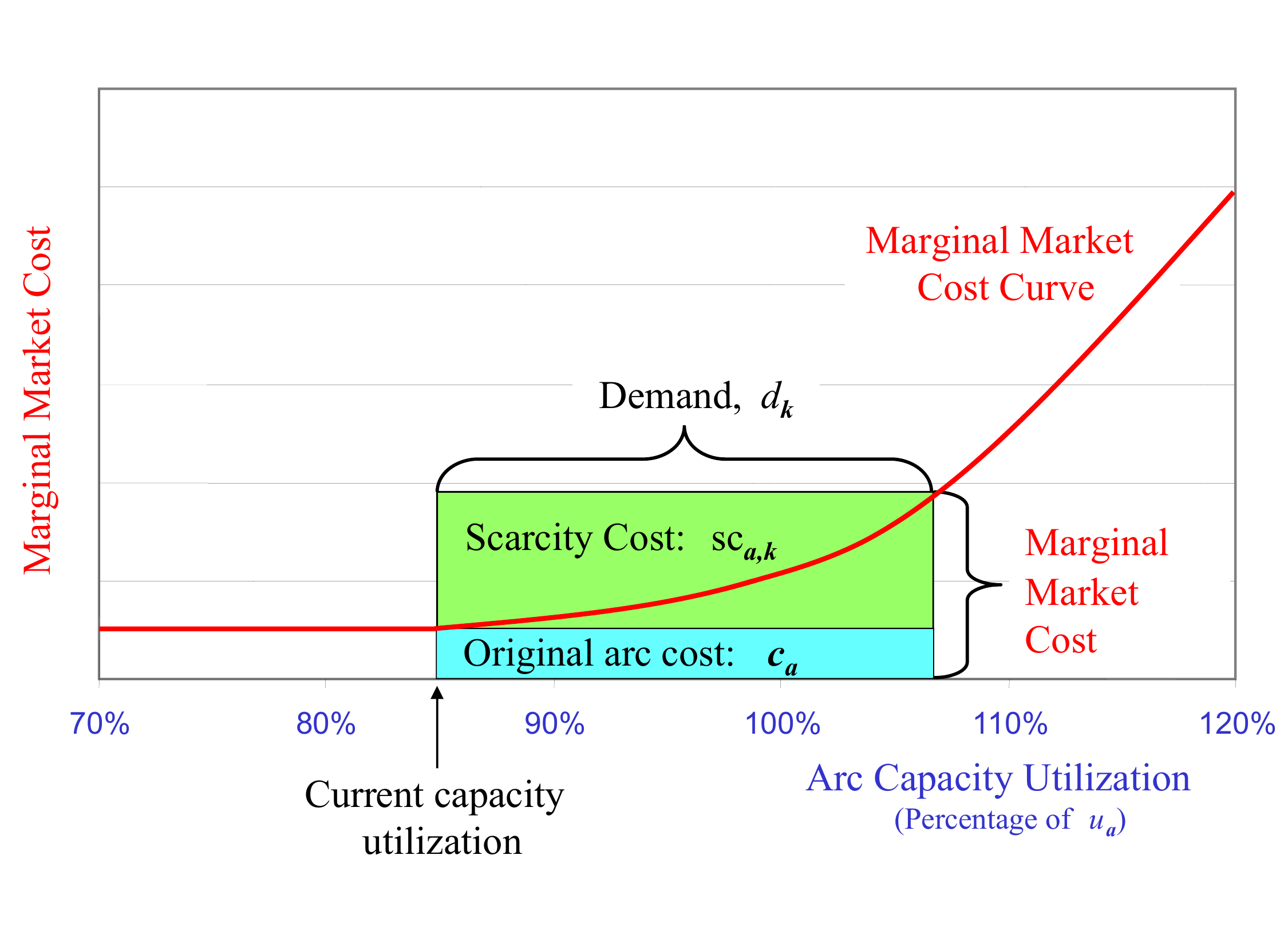}
	\end{center}
	\caption{Marginal Market Cost Curve} \label{mccurv} \end{figure}

\begin{figure}[tbp] \centering
     \includegraphics[angle = 0, scale = 0.4]{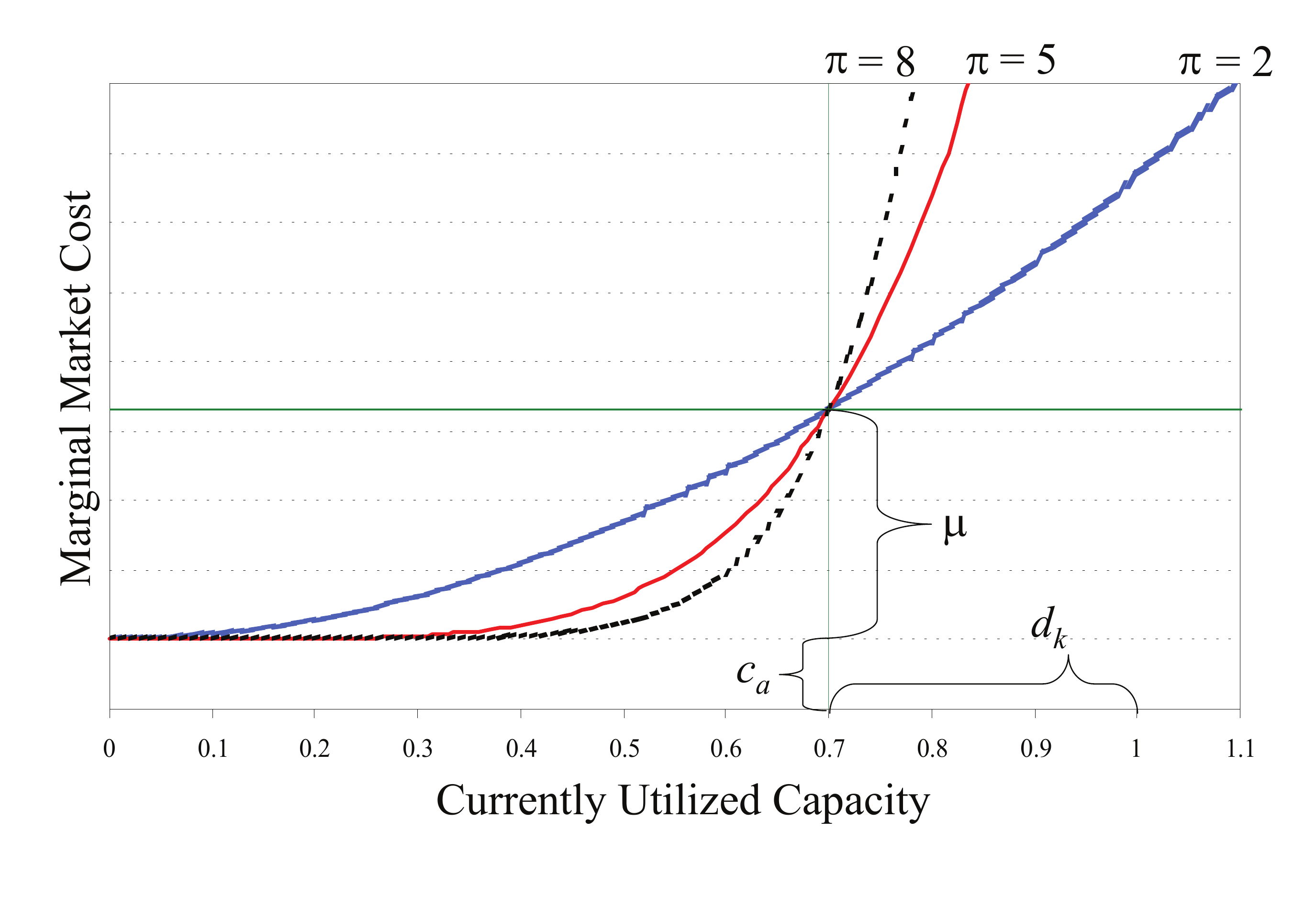}
    \caption{Marginal Market Cost Curve with Alternate $\pi$} \label{pcurv} \end{figure}

\begin{figure}[tb]
    \begin{center}
      \includegraphics[angle = 0, scale = 0.4]{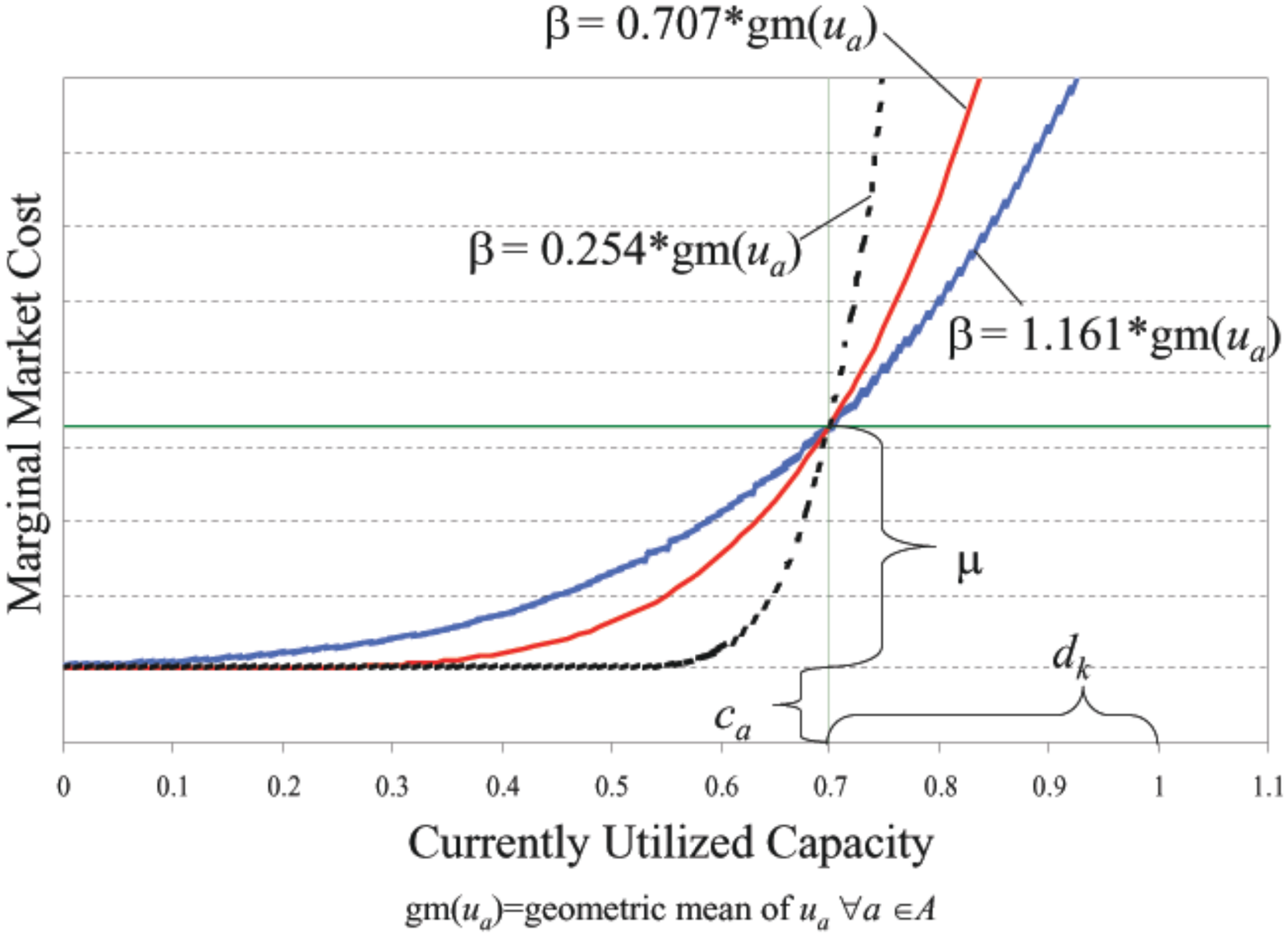}
    \end{center}
    \caption{Marginal Market Cost Curve with Alternate $\beta$} \label{bcurv} \end{figure}


\subsection{Scarcity Cost Parameters and the Market-Cost Curve}
\label{bpm}

The parameters $\beta$, $\mu$, and $\pi$ control the shape of the marginal
market-cost curve and determine at what arc capacity utilization the market cost is no longer infinitely elastic and the rate at which the market cost becomes increasingly inelastic. These effects are visible in the market-cost curve, 
the plot of market cost for arc $a \in A$ against currently
allocated arc capacity with respect to commodity $k \in K$, $u_a -
r(a,k, \mathbf X)$. Increasing  $\pi$ affects the rate of change
in the slope of the curve---how fast the price becomes inelastic. Increasing $\pi$ results in a decrease in marginal market cost for the region $ d_k < r(a,k, \mathbf X)$ and an
increase for the region $d_k > r(a,k, \mathbf X)$. The region
$d_k \le r(a,k, \mathbf X)$ corresponds to a set of flows for
which commodity $k$ can be routed on arc $a$ without violating the
capacity constraint for $a$, $u_a$. Decreasing $\pi$ has the
opposite affect. Figure \ref{pcurv} shows the marginal market-cost curve
with three different values of $\pi$ with all other parameters
held constant.

The parameter $\beta$ determines the point at which the scarcity
cost component of the market cost becomes non-zero and the marginal market cost is not infinitely elastic as $sc(a,k, \mathbf X)$
becomes non-zero when $r(a,k, \mathbf X) < \beta + d_k$. A secondary effect is that $\beta$ affects the slope of the curve as the scarcity
cost rises from 0 to $\mu$ over a change of $\beta$ in $(u_a -
r(a,k, \mathbf X))$. Figure \ref{bcurv} illustrates three
alternative values of $\beta$. 

Finally, parameter $\mu$ controls the
magnitude of $sc(a,k, \mathbf X)$ in a linear manner. Allocated
capacity values below zero are not shown for any cost curve as
$r(a,k, \mathbf X) \le u_a \Rightarrow (u_a - r(a,k, \mathbf
X)) \ge 0$. The parameters $\beta$, $\mu$, and $\pi$ may be used
to manipulate the shape of the market cost curve to adjust for
different applications. Section \ref{DE} describes one method for
adjusting the parameters for a specific application.

\subsection{IHH Algorithmic Steps}

The IHHO($\mathcal{P}$) heuristic for ODIMCF is given in Algorithm \ref{ihh}, where
$\mathcal{P} = ( N,A,K,\mathbf X$, $\beta$, $\mu$, $\pi)$
represents an ODIMCF problem, the current values for the decision
variables, and the values of the scarcity-cost parameters.

\begin{algorithm}
\begin{algorithmic}[1]

\REQUIRE $\mathcal{P}$

\ENSURE $\mathbf X$

\STATE $\mathbf X \leftarrow$SPSolve($\mathcal{P}$) \COMMENT{Route
commodities on minimum original arc cost paths.}

\IF[Is trivial SPSolve($\mathcal{P}$) solution feasible?]{
Feasible($A,K, \mathbf X$) = TRUE }

\STATE Stop

\ENDIF

\STATE $more \leftarrow \mathrm{TRUE}$ \COMMENT{Boolean variable
controlling termination of main loop.}

\STATE $\lambda_k \leftarrow 0 , \, \forall k \in K$ \COMMENT{Set
routing change counts to 0.}

\WHILE[Main loop.]{$more = \mathrm{TRUE}$}

\STATE $more \leftarrow \mathrm{FALSE}$

\STATE Randomize order of $K$

\FORALL[All commodities reexamine routing decision in random
order.]{$k \in K$}

 \IF[Does $k$'s routing decision change?]{ Route($k,  \lambda_k ,\mathcal{P}$) = TRUE }

\STATE $\lambda_k \leftarrow \lambda_k + 1$ \COMMENT{Increment
routing change counter.}

\STATE $more \leftarrow \mathrm{TRUE}$ \COMMENT{Market costs may
be altered for $g \in K \setminus \{k\}$.}

\ENDIF

\ENDFOR

\ENDWHILE

\STATE $\mathbf X \leftarrow$FeasPath($\mathcal{P}$) \COMMENT{Find
feasible routes with lower original arc costs.}

\STATE Return $\mathbf X$

\caption{IHHO($\mathcal{P}$)
Algorithm}\label{ihh}\end{algorithmic}\end{algorithm}

The
IHHO($\mathcal{P}$) algorithm starts with an initial solution
found by the SPSolve($\mathcal{P}$) procedure, which routes each commodity, $k$,
on the shortest path from $s_k$ to $t_k$ based on the original arc
costs without regard to arc capacities (Algorithm
\ref{spsolve}). This is done with the
SP($k,N,A$) algorithm for  the shortest-path problem defined
in (\ref{spb1})--(\ref{spb3}), where $\mathbf x$ is a vector of
flow variables for the shortest-path found. SP($k,N,A$)
 returns the set of
$s_k\rightarrow t_k$ path arcs or  $\emptyset$ if no path is found.

\begin{algorithm}
\begin{algorithmic}[1]

\REQUIRE $\mathcal{P}$

\ENSURE $\mathbf X$

\FORALL{ $k \in K$}

\STATE $NewPath \leftarrow$ SP($k,N,A$) \COMMENT{Find shortest
path from $s_k$ to $t_k$ using original arc costs.}

\IF[Does a path from $s_k$ to $t_k$ exist?]{$NewPath \ne
\emptyset$}

\STATE $\text{X}_{a,k} \leftarrow 1, \, \forall a \in NewPath$

\STATE $\text{X}_{a,k} \leftarrow 0, \, \forall a \in A \setminus
NewPath$

\ENDIF

\ENDFOR

\STATE Return $\mathbf X$ \caption{SPSolve($\mathcal{P}$)
Procedure}\label{spsolve}\end{algorithmic}\end{algorithm}

\begin{align}
&\Big[  \text{SP($k$)} \Big] \nonumber \\
&\text{Minimize:} & \sum_{a \in A} c_a x_a &\label{spb1}\\
&\text{subject to:} & \sum_{a \in T(n)} x_a - \sum_{a \in E(n)} x_a&= b^k_n &&\forall n \in N ,\, \forall k \in K \label{spb2}\\
&& x_a & \in \mathbb{B} && \forall a \in A \label{spb3}
\end{align}











This initial solution from SPSolve($\mathcal{P}$) is checked for feasibility with respect to the arc-capacity constraints by procedure Feasible($A,K,\mathbf X$) (not shown).  The solution is expected to be infeasible; if this trivial solution is feasible, IHHO($\mathcal{P}$) returns it as the optimal solution and exits.

After the initial solution is found in IHHO($\mathcal{P}$), the
variables $more$ and $\lambda_k$ are initialized. The boolean
variable $more$ controls the termination of the main loop. The
variable $\lambda_k \in \mathbb{Z}_{0+}$ is the count of routing
changes for commodity $k$ and is used to make a routing decision
in the Route($k, \lambda_k,\mathcal{P}$) procedure, described in the next section. IHHO($\mathcal{P}$) then iteratively reevaluates the
routing of each commodity until an equilibrium is reached and all
commodities are satisfied with their current routing decisions
based upon current marginal market costs. This state is indicated when
$more$ is FALSE.

During every iteration, each commodity reexamines its routing
 based upon current market costs, $mc(a,k, \mathbf X)$,
using the Route($k,  \lambda_k,\mathcal{P}$) procedure.  The order
in which commodities reevaluate their routing decisions is random.
Each commodity examines its decision once per iteration. This
ordering of commodities is implemented to avoid giving bias or
preferential treatment toward any single commodity or group of
commodities.  (If a preference for some
commodities is desirable, the ordering can be altered to reflect
that bias.)

A change in routing for commodity $k \in K$ is indicated by
the results of the Route($k,  \lambda_k,\mathcal{P}$) procedure: FALSE
(TRUE) indicates no change (a change). If no change occurred, the total flow and
 market costs for all arcs for all
commodities are also unchanged. If a reroute of $k$ is indicated, then the market costs for two or more arcs may have also changed for all other commodities.
Commodities having already evaluated their routing decisions
before $k$ during the current iteration of the main loop will
require the opportunity to reevaluate their decisions based on the
new market costs. This requirement is indicated by setting $more$
to TRUE and satisfied by executing a subsequent iteration.

The final step in IHHO($\mathcal{P}$) is the use of the
FeasPath($\mathcal{P}$) procedure, detailed in section \ref{feaspath} to search for lower-cost routes based on the original
arc costs.  This process only reassigns commodities to {\em
feasible routes}: routes with sufficient residual capacity.

\subsubsection{Routing Decision\label{routingdecision}}

Algorithm  \ref{route} shows the Route($k,\lambda_k,\mathcal{P}$)
procedure for reexamining the routing decision of commodity $k \in
K$ and adjusting the associated decision variables.  Let $CP(k) =
\{ a \in A: \text{X}_{a,k} = 1\}$ be the set of directed arcs
currently used by commodity $k \in K$. Commodity $k$ makes a
routing decision by considering the marginal market cost, the combination
of scarcity cost and original arc cost, in finding a route from
its origin, $s_k$, to its destination, $t_k$. 
Route($k,\lambda_k,\mathcal{P}$) determines a new route, $NewPath \subset A$,
based on market costs, which  is then compared with $CP(k)$, the
incumbent best route for commodity $k$, to determine if it is a new 
best route.

\begin{algorithm}
	\begin{algorithmic}[1]
		
		\REQUIRE $k \in K$, $\lambda_k \in \mathbb{Z}_{0+}$, $\mathcal{P}$
		
		\ENSURE TRUE if $k \in K$ has changed routing, else FALSE
		
		\STATE $NewPath \leftarrow \mathrm{SPS}(k,\mathcal{P})$
		\COMMENT{Find shortest $s_k-t_k$ path based on market costs.}
		
		\IF{$NewPath \ne \emptyset$}
		
		\IF[Has an improved new route been
		found?]{$rc(NewPath,k,A,\mathbf X) < hm(\lambda_k) rc(CP(k), k, A,
			\mathbf X)$}
		
		\STATE \COMMENT{Update decision variables.}
		
		\STATE $\text{X}_{a,k} \leftarrow 0 , \, \forall a \in CP(k)$
		
		\STATE $\text{X}_{a,k} \leftarrow 1 , \, \forall a \in NewPath$
		
		\STATE Return TRUE \COMMENT{Routing decision has changed.}
		
		\ENDIF \ENDIF
		
		\STATE Return FALSE \COMMENT{Routing decision remains the same.}
		
		\caption{Route($k$,$\lambda_k$,$\mathcal{P}$) Procedure}\label{route}\end{algorithmic}\end{algorithm}

$NewPath$ is found using SPS($k,\mathcal{P}$), which finds the
minimum cost path from $s_k$ to $t_k$ based on the current market cost for commodity $k \in
K$, $mc(a,k, \mathbf X)$. 
SPS($k,\mathcal{P}$) solves the problem in (\ref{sp_s1})--(\ref{sp_s3}),
where $\mathbf x$ is a vector of binary flow variables determining
$NewPath$. SPS($k,\mathcal{P}$) returns the set of arcs in the
shortest-path discovered or  $\emptyset$ if no path is found.

\begin{align}
 &\Big[ \text{SPS($k$)} \Big] \nonumber \\
 &\text{Minimize:}& \sum_{a \in A} mc(a,k, \mathbf X) x_a &\label{sp_s1}\\
 &\text{subject to:}& \sum_{a \in T(n)} x_a - \sum_{a \in E(n)} x_a &= b^k_n && \forall n \in N \label{sp_s2}\\
 &&  x_a &\in \mathbb{B} && \forall a \in A \label{sp_s3}
\end{align}

The value of the variable $\lambda_k$ passed to Route($k, \lambda_k,
\mathcal{P}$) by IHHO($\mathcal{P}$) is the number of times
commodity $k \in K$ has changed routing. Let the cost hurdle
multiplier, $hm(\lambda_k) \rightarrow \mathbb{R}$, be a (user-defined) monotonically decreasing function in the range [0,1] such
that $\lambda_k \in \mathbb{Z}_{0+}$ and $\exists \lambda_0 < \infty
\text{ s.t. } hm(\lambda_0) = 0$. The new route, $NewPath$,
replaces the incumbent route, $CP(k)$, only if
$rc(NewPath,k,A,\mathbf X) < hm(\lambda_k)
rc(CP(k),k,A,\mathbf X)$. $NewPath$ must provide a certain level
of improvement over the incumbent with respect to current market
costs for a replacement to occur. As commodity $k$ changes routes
more regularly, $\lambda_k$ will increase and $hm(\lambda_k)$ will
decrease, as shown in Figure \ref{mccurv}. The gradual decrease in the cost hurdle multiplier requires subsequent $NewPath$s to provide an increasingly
substantial improvement over $CP(k)$. 
To allow IHHO(P) to achieve an equilibrium without being artificially forced into the equilibrium, $hm(\lambda_k)$ is designed to have a value of 1 until $\lambda_k \geq \lambda_1$. With $hm(\lambda_k)=1$, routes $NewPath$ and $CP(k)$ are compared based solely on their marginal market costs.

\begin{figure}[tbp]
	\begin{center}
		\includegraphics[scale=0.5]{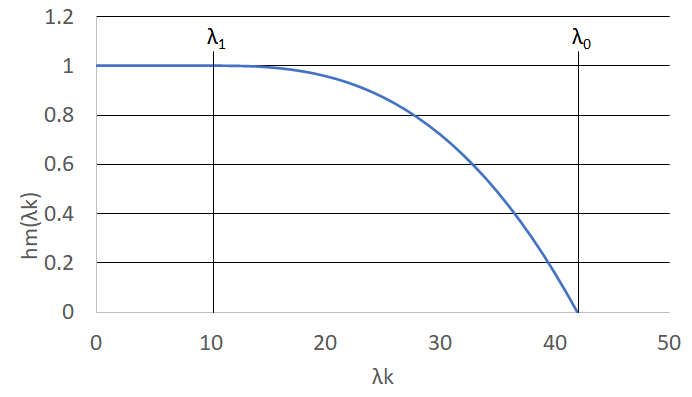}
	\end{center}
	\caption{Hurdle Multiplier Curve for $\lambda_1$ and $\lambda_0$.} \label{mccurv} \end{figure}

With other commodities changing routes, $rc(CP(k),k,A,\mathbf
X)$ may change between iterations even if $CP(k)$ is the same.  If
this path's residual capacity decreases  significantly and the
current $CP(k)$ becomes untenable,
$hm(\lambda_k)rc(CP(k),k,A,\mathbf X)$ will increase, possibly
enabling a previously rejected route to replace the incumbent.

Creating $hm(\lambda_k)$ such that $hm(\lambda_0) = 0$ for some
$\lambda_0 < \infty$ ensures each commodity can change routes at
most $\lambda_0$ times, as no $NewPath$ can have a cost less than 0. If IHHO($\mathcal{P}$) is unable to reach an equilibrium, this limit on the number of changes per commodity ensures the termination of IHHO($\mathcal{P}$).

\subsubsection{FeasPath($\mathcal{P}$) Procedure\label{feaspath}}

The marginal market cost curve used in the main loop of the IHHO(P) algorithm is not a precise match to the actual objective function and constraints in the original ODIMCF problem. For an arc with capacity utilization approaching its capacity constraint, the marginal market cost for a commodity may greatly exceed the original arc cost even if the arc has sufficient residual capacity to route the commodity without violating the arc's capacity constraint. This cost mismatch may encourage some commodities to choose a path with a lower marginal market cost but with a higher original cost in the ODIMCF problem. Conversely, the marginal market cost curve allows for an arc's utilization to exceed an arc's capacity but at a steep marginal market cost. Exceeding the arc's capacity is permitted to determine the demand for a specific arc to enable the rationing function of the demand curve. While it is expected---and testing indicates (Section \ref{testresults})---the main loop in IHHO($\mathcal{P}$) finds a solution close to the original ODIMCF problem, a final procedure, FeasPath($\mathcal{P}$) attempts to close any remaining gaps by changing the supply curve to be infinitely elastic up to full arc capacity utilization and infinitely inelastic once an arc's capacity is fully utilized.

The FeasPath($\mathcal{P}$) procedure, Algorithm \ref{feaspathalgo},
examines the commodities to find routes incurring lower original-arc cost while not violating arc capacity constraints. If an
improved route is found, $CP(k)$ is switched to the new path. If
an improved route is not found, $CP(k)$ is left as is. If $CP(k)$
is not altered by FeasPath($\mathcal{P}$), $CP(k)$ may contain
arcs whose capacity constraints are violated.
FeasPath($\mathcal{P}$) enforces capacity constraints only for
altered routes. FeasPath($\mathcal{P}$) does not un-route routed
commodities; $CP(k) = \emptyset$ after FeasPath($\mathcal{P}$)
only if $CP(k) = \emptyset$ at the start of
FeasPath($\mathcal{P}$). 

\begin{algorithm}
	\begin{algorithmic}[1]
		
		\REQUIRE $\mathcal{P}$
		
		\ENSURE $\mathbf X$
		
		\STATE $\lambda_k \leftarrow 0 \mathrm{, } \forall k \in K$
		
		\STATE $more \leftarrow \mathrm{TRUE}$
		
		\WHILE{$more = \mathrm{TRUE}$}
		
		\STATE $more \leftarrow \mathrm{FALSE}$
		
		\STATE Randomize order of $K$
		
		\FORALL{$k \in K$}
		
		\STATE $NewPath \leftarrow $ SPF($k,\mathcal{P}$)
		
		\IF{$frc(NewPath, k,K \mathbf X) < min(\text{M},hm(\lambda_k) frc(CP(k), k, \mathbf X))$}
		
		\STATE $\lambda_k \leftarrow \lambda_k + 1$
		
		\STATE $\text{X}_{a,k} \leftarrow 0 , \, \forall a \in CP(k)$
		
		\STATE $\text{X}_{a,k} \leftarrow 1 , \, \forall a \in NewPath$
		
		\STATE $more \leftarrow \mathrm{TRUE}$
		
		\ENDIF
		
		\ENDFOR
		
		\ENDWHILE
		
		\STATE Return $\mathbf X$
		
		\caption{FeasPath($\mathcal{P}$)
			Procedure}\label{feaspathalgo}\end{algorithmic}\end{algorithm}


The {\em feasible arc cost}, $fc(a,k,K,\mathbf X) \rightarrow \mathbb{R}_{0+}$ (\ref{fc}), is given as the original arc cost, $c_a$,  if
arc $a$ has at least $d_k$ residual capacity, otherwise 
 infinity. These are used to determine the {\em feasible route cost} of route $P$ for commodity $k$, $frc(P,k,\mathbf X) \rightarrow \mathbb{R}_{0+}$  (\ref{frc}).

\begin{align} fc(a,k,K,\mathbf X) &= \begin{cases}
c_a &\text{if $r(a,k, \mathbf X) - d_k \ge 0$}, \\
\text{M} & \text{otherwise}.
\end{cases}  \label{fc} \\
frc(P,k,\mathbf X) & = \sum_{ a \in P} fc(a,k,K,\mathbf X)\label{frc}
\end{align}

\noindent where scalar $\text{M} \ge max(\mathbf c)|N|$.

\begin{align}
 &\Big[  \text{SPF($k$)} \Big] \nonumber \\
 &\text{Minimize:}&  \sum_{a \in A} fc(a,k,K,\mathbf X) x_a  &\label{sp_f1}\\
  &\text{subject to:}& \sum_{a \in T(n)} x_a - \sum_{a \in E(n)} x_a  &= b^k_n && \forall n \in N \label{sp_f2}\\
  && x_a &\in  \mathbb{B} && \forall a \in A \label{sp_f3}
\end{align}

FeasPath($\mathcal{P}$) solves SPF($k,\mathcal{P}$) for each
commodity $k \in K$ having its routing reexamined. The formulation
for the problem solved is shown in equations (\ref{sp_f1})--(\ref{sp_f3}). SPF($k,\mathcal{P}$) returns the set of arcs in the
shortest path based on feasible arc costs from $s_k$ to $t_k$. If
$frc(NewPath,k,\mathbf X) < \text{M}$ then this is the
shortest-path with respect to original arc cost on which
sufficient residual arc capacity exists to route commodity $k$. If
$frc(NewPath,k, \mathbf X) \ge \text{M}$ then there does not
exist a path from $s_k$ to $t_k$ with at least $d_k$ residual
capacity on each arc and no feasible path exists.

\subsection{Interpretation of IHH's Final Solution}

The final solution found by IHHO($\mathcal{P}$) provides a single
route for each commodity through the network and will always meet
the ODIMCF node-balance and integrality constraints (\ref{odb2}) and (\ref{odb4}). If the final solution
is feasible with respect to the arc capacity constraints
(\ref{odb3}), then an integer feasible solution is at
hand. Unlike some integer programming techniques,
IHHO($\mathcal{P}$) does not provide a bound on the optimality
gap, the difference between the objective function values of the
optimal solution and the IHHO($\mathcal{P}$) solution. Other
techniques often rely on the linear programming relaxation as a
source of gap information. While this relaxation of ODIMCF can be
solved to provide such gap information, testing indicates the time
required to solve the relaxation will be greater than the time
required by IHHO($\mathcal{P}$) to find an integer feasible
solution.

If the solution found by IHHO($\mathcal{P}$) violates one
or more arc capacities, the solution will be infeasible, meaning no feasible solution to that particular ODIMCF problem exists or the heuristic
could not identify such a solution. This
situation was not encountered in the computational testing.

\subsection{Asymptotic Bounds\label{asymptotic}}

IHHO($\mathcal{P}$) has polynomial asymptotic bounds with respect
to both time and space.  The asymptotic bound on space
requirements is O($|A| + |N||K|$). The $|A|$ term represents the
space needed to store arc information. The $|N||K|$ term
represents the space required to store route information.  As the
assumption is made that arc costs are always non-negative, a route
will contain at most $|N| - 1$ arcs with no cycles.  One route is
stored for each commodity. The O($|N|$) space required for storage
of node information is dominated by the $|N||K|$ term. Similarly,
a $|K|$ term representing commodity information is omitted.

The asymptotic bound on running time for IHHO($\mathcal{P}$) is
$\text{O}(\lambda_0|K|^2(|A| + |N|\log|N|))$. SP($k,N,A$),
SPF($k,\mathcal{P}$), and SPS($k,\mathcal{P}$) use a shortest-path
algorithm with a time bound of O($|A| + |N|\log|N|$), under the
assumption of non-negative arc costs \cite{AHUJ93}.
Route($k,\lambda_k,\mathcal{P}$) uses SPS($k,\mathcal{P}$) once and
requires O($|N|$) additional time to update and compare routes for
a time bound of O($|A|$ + $|N|\log|N|$). SPSolve($\mathcal{P}$) uses
SP($k,N,A$) and records a route once for each commodity for a
total of O($|K|(|A| + |N|\log|N|)$) time. Each commodity can change
routes at most $\lambda_0$ times before the cost-hurdle makes
further change impossible.  In the worst case, at most one
commodity will change routes during each iteration of the main
loop of IHHO($\mathcal{P}$). This worst case results in
$\lambda_0|K|$ executions of the main loop requiring
O($\lambda_0|K|^2(|A| + |N|\log|N|)$) time. The main loop within
FeasPath($\mathcal{P}$) is similar to the main loop of
IHHO($\mathcal{P}$) resulting in a O($\lambda_0|K|^2(|A| +
|N|\log|N|)$) time bound for FeasPath($\mathcal{P}$).

\section{Computational Testing}

Computational testing is designed to determine the performance
characteristics of an IHHO($\mathcal{P}$) implementation and compare them to commercial-grade software. Test
problems are generated to measure the responses of running time
and solution quality to changes in network and commodity
characteristics. 

The reported running times do not include time to
read the problem or record the solutions. Solution quality
is compared with values for the LP relaxation (LPR) and the
best-available integer programming (MIP) solution.

\subsection{Test Environment}

All benchmark testing is performed on a Dell R720 with dual Dual Six Core Intel Xeon 3.5GHz	processors and 252GB RAM at Southern Methodist University's Lyle School of Engineering. The IHHO($\mathcal{P}$) algorithm is implemented in C++
and compiled with {\tt g++} at the
default optimization level.  Reported running times (CPU
execution times) are exclusive of input and output processing.

LPR and MIP solutions are generated
using IBM ILOG CPLEX Interactive Optimizer 12.6.0.0 (CPLEX12).  CPLEX12 is run with default settings, with three exceptions: the MIP time limit is
set to 7200 seconds, the optimality tolerance increased to
0.25\%, and single-thread mode was used.\footnote{This time limit is set to ensure a timely termination
of testing. The optimality tolerance is increased as initial
testing revealed CPLEX12 would expend a large amount of effort
closing the optimality gap after finding a  good, even optimal,
solution. Since our implementation has not been designed for multiple threads, 
the single-thread mode for CPLEX12 was used for comparability.}

\subsection{Parameter Selection} \label{DE}

Performance of IHHO($\mathcal{P}$) is affected by the parameters
$\beta$, $\mu$, and $\pi$.  To avoid manually varying these values to find a set of parameters with good performance over a range of problems, the metaheuristic differential evolution (DE) \cite{CORN99} was used to select their
values, as follows. Three problems were
generated for four groups with different problem dimensions. 
DE is an evolutionary algorithm with a population. For this application, a member of the population was a defined by 3-tuple of values for 
$\beta$, $\mu$, and $\pi$. DE evaluated each member of the population by running the problems through IHHO(P) with the member's $\beta$, $\mu$, and $\pi$ values and taking the mean of the routing cost normalized to known solution values.  Members evaluated with lower cost were preferred during the creation of the next generation. A population size of 30 was used.  Following 100 DE generations, the benchmark testing values for
$\beta$, $\mu$, and $\pi$ were determined from the group results
and are shown in Table \ref{params}.
	The parameters were tuned on a completely different set of problems from those 
	used in the testing reported herein. In addition, the two sets 
	of problems were generated using different problem generators for
	both the networks and commodities.

\begin{table}[tbp]
	\begin{center}
		\caption{IHHO($\mathcal{P}$) Parameters}

		\begin{tabular}{ll}
			Parameter & Value \\
			\hline
			$\beta$ & $ max\Big( \frac{\sqrt{2}}{2}gm(u_a) \mathrm{, } gm({d_k}) \Big)$ \\
			$\mu$& $ gm(c_a) \sqrt{e}$ \\
			$\pi$ & $e^e$ \\
			$\lambda_0$ & 43 \\ 
			$\lambda_1$ & 10 \\
			$hm(\lambda_k)$ & 1, if $\lambda_k <\lambda_1$; else $1-\left(\frac{\lambda_k-\lambda_1}{\lambda_0-\lambda_1-1}\right)^e$ \\ \hline
			
			\multicolumn{2}{l}{ \footnotesize{$gm(y_z)$ is the geometric mean
					of values for $y_z$ $\forall z \in Z$.}}
		\end{tabular}
		\label{params} \end{center} \end{table}

While DE used IHHO(P) as a subroutine to automatically determine a set of parameters, IHHO($\mathcal{P}$) has no dependency on DE. Other methods for tuning the parameters $\beta$, $\mu$, and $\pi$ could have been used. If IHHO($\mathcal{P}$) is to be used for a new application and similar benchmark problems are available, re-tuning the parameters is recommended. The use of an automated tuning method such as DE would facilitate periodic reevaluation of these parameters, and new problems could be added to the set of benchmark problems to optimize against.

Table 4 contains the expression used for the hurdle multiplier hm($\lambda_k$) and its parameters $\lambda_0$ and $\lambda_1$. While the hurdle multiplier is used to guarantee termination of IHHO($\mathcal{P}$), the algorithm converged to an equilibrium state before the hurdle multiplier had any effect for almost all problems as described in section \ref{testresults}. Therefore, no effort was made to tune $\lambda_0$ and $\lambda_1$ and the initially selected values were used for all testing.

\subsection{Problem Generator}

To explore the effects of underlying network topology and other
problem characteristics, an ODIMCF problem generator, ODGEN,\footnote{Source code  available from the authors upon request.} was
developed.  ODGEN accepts as input parameters the number of
commodities, the number of nodes, number of directed arcs, arc
cost range (given as a minimum and 90th percentile value), and
commodity demand range (given as a minimum and maximum value).

ODGEN initially assigns a random position to each node in a
two-dimensional space before determining the set of arcs. Arcs are
generated to form a {\em mesh} network topology by having the probability
of an arc connecting $i \in N$ to $j \in N\setminus \{i\}$ be
inversely proportional to the distance in the two-dimensional
space from $i$ to $j$ raised to a certain power. Once set $A$ is
generated, the arc costs are set to the arc distance in the
two-dimensional space scaled so that the minimum and 90th
percentile arc costs match the user input. All networks are 
connected graphs in that a path exists from every node to every other
node.\footnote{For industry problems with multiple subgraphs that are not connected to each other, a solution algorithm 
should be  applied to each such subgraph separately.}  For every arc $a \in A$ there does not exist a path from
$i_a$ to $j_a$ with a cost less than $c_a$.  The networks do not
contain parallel arcs.

For each commodity $k \in K$, $s_k$ is randomly chosen from $N$
with each node having the same probability of being chosen and
$t_k$ is similarly selected from $N \setminus \{s_k\}$. The demand
$d_k$ is chosen randomly between the minimum and maximum using a
uniform distribution.

After the commodities are generated, each commodity is routed on a
{\em random-shortest path} from $s_k$ to $t_k$ found as the
shortest path with the arc lengths set randomly between 1 to
10,000 and changing for each commodity. The arc capacities problem
are set to the sum of the capacity required by the commodities
using these random-shortest paths. This method of setting arc
capacities ensures that a feasible solution exists, but not
necessarily using the shortest-paths based on the arc costs in the
problem.

To enable experimentation of non-uniform distribution of origins and destination, the generator allows for the specification of a percentage of vertices to be designated as hubs. A specified percentage of commodities must be routed between an origin hub and a destination hub so that the bulk of the flow will be between hubs possibly passing through other vertices in the mesh network. The remaining commodities are generated as described previously still allowing for the hub vertices to be paired with non-hub vertices. The arc capacities are determined as previously described to ensure a feasible solution exists.

\subsection{Test Problems' Characteristics\label{problemCharacteristics}}

Three test sets A, H, and L are created to evaluate the effects of $|N|, |A|$, and $|K|$ on IHHO(P)'s performance and to explore the method's ability to solve much larger instances than previously published. Sets A and H have the same network topologies with set A having commodity origins and destinations uniformly distributed as with a mesh network structure; set H having certain vertices acting as hubs, as found in logistics and distribution networks with higher interactions between the hub nodes. Test set L is similar to set A with uniform origin and destination distribution for commodities, but with significantly larger values for $|N|, |A|$, and $|K|$ to analyze IHHO($\mathcal{P}$)'s runtime performance for increasingly large problem sets. All problems are known to have feasible solutions with respect to arc capacity constraints.

The characteristics of all three test problem sets are shown in Table \ref{agc}. Within sets A and H eight groups of different problems with similar characteristics (number of nodes, arcs, commodities, average commodity demand, mean arc capacity, and average node degree) are created. Within each group, five different test problems are randomly generated with identical values for $|N|, |A|$, and $|K|$, but with different random-number seeds. Arc costs are set with a minimum value of 10 and a value of 2000 for the 90th percentile. Commodity demands range from 5 to 25. Arc capacities are determined by ODGEN to ensure feasibility. For set H, 10 percent of vertices are hubs and 80 percent of commodities must have hub vertices as both origin and destination. The large problem set L contains six groups, also shown in Table \ref{agc}. Within each group, five different test problems are randomly generated with identical values for $|N|, |A|$, and $|K|$ but with different random-number seeds.

\begin{sidewaystable}[tbp]
	\caption{Test Problem Sets:  Group Characteristics  \label{agc} }
	\begin{tabular} {lrrrrrrrrrcc}
		\mbox{\rule{0pt}{20pt}} Group & \multicolumn{1}{c}{$|N|$} &
		\multicolumn{1}{c}{$|A|$} & \multicolumn{1}{c}{$|K|$} &
		\multicolumn{1}{c}{0/1 variables} &
		\multicolumn{1}{c}{Constraints} &
		\multicolumn{1}{c}{$\overline{ c_a}$} &
		\multicolumn{1}{c}{$\overline {d_k}^1$} &
		\multicolumn{1}{c}{$\overline{ u_a}^2$} &
		\multicolumn{1}{c}{$\overline{ degree}$} &
		\multicolumn{1}{c}{$\overline{\left( \frac{\sum_{k \in K} d_k }{
					\sum_{a \in A} u_a}\right)}^3$}
		& \multicolumn{1}{c}{$\overline{\left(\frac{ \overline{d_k}}{\overline{ u_a} }\right)}^4$ }\\
		\hline
		A1                 & 30&  90&  112&10,196 & 3,454 & 	2,331&	15.18&  122.5&6&0.15&0.12\\
		A2                 & 30&  90&  281& 25,575 & 8,524 & 	2,331&	15.13&  248.4&6& 0.19&0.06\\
		A3                 & 30& 360& 1,728& 623,812 & 52,204 & 2,372&		15.09& 161.6&24& 0.45&0.09\\
		A4                 & 30& 360& 4,320& 1,559,524 & 129,964 & 		2,372&15.03& 384.5&24& 0.47&0.04\\ 
		A5                 &120& 360&  257& 92.81 & 31,204 & 	1,968&	15.50&  149.7&6&0.07&0.10\\
		A6                &120& 360&  642& 231,766 & 77,404 & 	1,968&	15.15&  312.2&6& 0.09&0.05\\
		A7                &120&1,440& 4,937& 7,114,221 & 593,884 & 		1,331&14.96& 195.1?&24& 0.26&0.08\\
		A8                &120&1,440&12,342& 17,784,826 & 1,482,484 & 		1,331&14.98& 447.5&24& 0.29&0.03\\
		\hline
		H1                 & 30&  90&  112&10,196 & 3,454 & 			 2,331 	&	 15.18 	&	 149.7 	&	 6 	&	0.64	&	 0.10 	\\
		H2                 & 30&  90&  281& 25,575 & 8,524 & 		 2,331 	&	 15.13 	&	 283.4 	&	 6 	&	0.74	&	 0.05 	\\
		H3                 & 30& 360& 1,728& 623,812 & 52,204 & 		 2,372 	&	 15.09 	&	 207.2 	&	 24 	&	0.63	&	 0.07 	\\
		H4                 & 30& 360& 4,320& 1,559,524 & 129,964 & 		 2,372 	&	 15.03 	&	 388.7 	&	 24 	&	0.74	&	 0.04 	\\ 
		H5                 &120& 360&  257& 92.81 & 31,204 & 		   1,968 	&	 15.50 	&	 152.9 	&	 6 	&	0.61	&	 0.10 	\\
		H6                &120& 360&  642& 231,766 & 77,404 & 		 1,968 	&	 15.15 	&	 281.9 	&	 6 	&	0.75	&	 0.05 	\\
		H7                &120&1,440& 4,937& 7,114,221 & 593,884 & 		1,331 	&	 14.96 	&	 199.3 	&	 24 	&	0.71	&	 0.08 	\\
		H8                &120&1,440&12,342& 17,784,826 & 1,482,484 & 		 1,331 	&	 14.98 	&	 448.3 	&	 24 	&	0.75	&	 0.03 	\\
		\hline
L1	&	 480 	&	 5,760 	&	 15,360 	&	            88,473,600  	&	            7,378,560  	&	 1,883.5 	&	 15.0 	&	 371.9 	&	24	&	 0.11 	&	 0.04 	 \\
L2	&	 480 	&	 5,760 	&	 38,400 	&	          221,184,000  	&	          18,437,760  	&	 1,883.5 	&	 15.0 	&	 913.5 	&	24	&	 0.11 	&	 0.02 	 \\
L3	&	 960 	&	 11,520 	&	 25,134 	&	          289,543,680  	&	          24,140,160  	&	 2,159.5 	&	 15.0 	&	 366.1 	&	24	&	 0.09 	&	 0.04 	 \\
L4	&	 960 	&	 11,520 	&	 62,836 	&	          723,870,720  	&	          60,334,080  	&	 2,159.5 	&	 15.0 	&	 895.7 	&	24	&	 0.09 	&	 0.02 	 \\
L5	&	 1,920 	&	 23,040 	&	 42,535 	&	          980,006,400  	&	          81,690,240  	&	 3,383.2 	&	 15.0 	&	 367.5 	&	24	&	 0.08 	&	 0.04 	 \\
L6	&	 1,920 	&	 23,040 	&	 106,338 	&	      2,450,027,520  	&	        204,192,000  	&	 3,383.2 	&	 15.0 	&	 902.4 	&	24	&	 0.08 	&	 0.02 	 \\
\hline
		\multicolumn{10}{l}{\footnotesize $^1$Average demand per commodity}\\
		\multicolumn{10}{l}{\footnotesize $^2$Mean arc capacity}\\
		\multicolumn{10}{l}{\footnotesize $^3$Average ratio of total routed demand to combined capacity of all network arcs}\\
		\multicolumn{10}{l}{\footnotesize $^4$Average ratio of mean destination demand to mean arc capacity}
	\end{tabular}
\end{sidewaystable}

\subsection{Test Set Results\label{testresults}}

Table \ref{probtimesA} shows the test problem run times and final solution costs for
the IHH code and CPLEX's LP relaxation and integer programming solvers for Test Set A.
Solution times are in CPU seconds and the IHH problem times are an average of  ten
runs with different random number seeds. Since  IHHO($\mathcal{P}$)  randomizes the commodity consideration 
order, ten random-number-generator seeds are used to solve each
problem instance. Each reported group's results represent 50 combinations of
problem and seed.)

Table \ref{probtimesA} provides computational results for the test set A.   IHHO($\mathcal{P}$) found feasible
solutions to all problems and CPLEX did not find a feasible MIP solution for eleven problems and one LPR  problem  within the time limit.
The table also provides the ratios in objective function values between IHH and the LPR (linear programming relaxation) and the MIP (mixed integer programming) solution values provided by CPLEX12. IHH costs averaged 3.5\% higher (median 2.4\%) than the non-integer LPR solutions and a mean of 3.1\% (median 2.0\%) above MIP solution values. But these high-quality IHH solutions were identified in a fraction of the time required by CPLEX12, as shown later.

\begin{table}[tbp]
	\begin{center}
		\caption{ Test Set A Problems' Solution Times and Costs\label{probtimesA}}
		{\footnotesize
			\begin{tabular} {lcrrrrrrrr}
				
				&		&	 \multicolumn{3}{c}{Solution times, in seconds} 					&	 \multicolumn{3}{c}{Solution cost} & \multicolumn{2}{c}{IHH ratio$^2$ to:} 					\\
				Grp	&	Prob	&	 IHH$^1$ 	&	 LPR 	&	 MIP 	&	 IHH$^1$ 	&	 LPR 	&	 MIP&LPR&MIP 	\\ \hline
A1	&	A1	&	0	&	0.09	&	1.27	&	25,888,055	&	25,219,739	&	25,380,948	&	 1.026 	 & 	 1.020 	\\	
&	A2	&	0	&	0.04	&	0.45	&	13,767,055	&	13,619,834	&	13,683,181	&	 1.011 	 & 	 1.006 	\\	
&	A3	&	0	&	0.06	&	0.36	&	15,409,031	&	14,307,332	&	14,433,630	&	 1.077 	 & 	 1.068 	\\	
&	A4	&	0.1	&	0.11	&	5.47	&	17,565,724	&	16,979,419	&	17,091,778	&	 1.035 	 & 	 1.028 	\\	
&	A5	&	0	&	0.06	&	1.41	&	5,999,320	&	5,829,935	&	5,851,626	&	 1.029 	 & 	 1.025 	\\	\\
A2	&	A11	&	0.01	&	0.27	&	4.12	&	65,433,671	&	64,183,982	&	64,382,083	&	 1.019 	 & 	 1.016 	\\	
&	A12	&	0.01	&	30.68	&	0.81	&	32,559,908	&	32,134,593	&	32,210,139	&	 1.013 	 & 	 1.011 	\\	
&	A13	&	0.01	&	0.14	&	0.88	&	35,452,909	&	34,619,069	&	34,723,468	&	 1.024 	 & 	 1.021 	\\	
&	A14	&	0.02	&	0.1	&	0.93	&	45,972,645	&	45,424,869	&	45,498,683	&	 1.012 	 & 	 1.010 	\\	
&	A15	&	0.02	&	0.09	&	3.34	&	14,470,213	&	14,211,330	&	14,238,794	&	 1.018 	 & 	 1.016 	\\	\\
A3	&	A21	&	0.9	&	29.8	&	931.37	&	139,968,253	&	131,959,935	&	132,849,007	&	 1.061 	 & 	 1.054 	\\	
&	A22	&	0.68	&	12.65	&	839.59	&	64,932,431	&	61,226,836	&	61,846,681	&	 1.061 	 & 	 1.050 	\\	
&	A23	&	0.71	&	4.97	&	329.55	&	78,437,815	&	74,377,380	&	74,804,983	&	 1.055 	 & 	 1.049 	\\	
&	A24	&	0.75	&	10.6	&	854.71	&	95,362,616	&	90,247,267	&	90,659,190	&	 1.057 	 & 	 1.052 	\\	
&	A25	&	0.66	&	17.34	&	2,144.00	&	107,652,301	&	100,924,979	&	101,866,634	&	 1.067 	 & 	 1.057 	\\	\\
A4	&	A31	&	2.01	&	153.18	&	1,110.44	&	328,010,021	&	320,334,827	&	$\dagger$	&	 1.024 	 & 	$\dagger$	\\	
&	A32	&	1.81	&	91.07	&	1,110.44	&	154,170,274	&	150,964,798	&	151,805,158	&	 1.021 	 & 	 1.016 	\\	
&	A33	&	1.73	&	86.92	&	885.83	&	186,592,310	&	183,373,456	&	184,218,518	&	 1.018 	 & 	 1.013 	\\	
&	A34	&	1.71	&	53.79	&	6,578.58	&	228,166,279	&	223,715,157	&	225,323,285	&	 1.020 	 & 	 1.013 	\\	
&	A35	&	2.21	&	162.12	&	1,493.71	&	250,631,458	&	245,005,840	&	246,759,411	&	 1.023 	 & 	 1.016 	\\	\\
A5	&	A41	&	0.07	&	2.4	&	11.37	&	75,910,921	&	70,833,506	&	71,216,145	&	 1.072 	 & 	 1.066 	\\	
&	A42	&	0.06	&	0.78	&	4.04	&	75,772,046	&	72,847,682	&	73,275,906	&	 1.040 	 & 	 1.034 	\\	
&	A43	&	0.05	&	0.44	&	5.48	&	52,370,346	&	49,806,807	&	50,229,531	&	 1.051 	 & 	 1.043 	\\	
&	A44	&	0.06	&	1.01	&	5.34	&	73,467,522	&	68,774,329	&	69,031,890	&	 1.068 	 & 	 1.064 	\\	
&	A45	&	0.06	&	0.95	&	79.36	&	58,605,039	&	53,891,427	&	54,589,672	&	 1.087 	 & 	 1.074 	\\	\\
A6	&	A51	&	0.17	&	15.73	&	68.43	&	174,513,790	&	170,467,198	&	171,025,999	&	 1.024 	 & 	 1.020 	\\	
&	A52	&	0.19	&	4.18	&	41.6	&	179,542,511	&	177,252,846	&	177,679,960	&	 1.013 	 & 	 1.010 	\\	
&	A53	&	0.16	&	2.66	&	10.04	&	119,258,159	&	118,218,052	&	118,436,013	&	 1.009 	 & 	 1.007 	\\	
&	A54	&	0.18	&	7.34	&	36.53	&	171,600,499	&	168,819,235	&	169,346,548	&	 1.016 	 & 	 1.013 	\\	
&	A55	&	0.21	&	4.76	&	72.07	&	130,883,159	&	128,197,322	&	128,819,852	&	 1.021 	 & 	 1.016 	\\	\\
A7	&	A61	&	10.25	&	4,073.79	&	7,201.14	&	163,207,827	&	156,309,412	&	$\dagger$	&	 1.044 	 & 	$\dagger$	\\	
&	A62	&	10.71	&	1,894.73	&	7,200.50	&	300,256,560	&	290,068,077	&	$\dagger$	&	 1.035 	 & 	$\dagger$	\\	
&	A63	&	9.8	&	1,881.49	&	7,200.91	&	338,067,201	&	323,369,229	&	$\dagger$	&	 1.045 	 & 	$\dagger$	\\	
&	A64	&	11.58	&	2,211.85	&	7,201.62	&	197,008,935	&	187,343,861	&	$\dagger$	&	 1.052 	 & 	$\dagger$	\\	
&	A65	&	12.96	&	4,731.51	&	7,471.79	&	255,714,681	&	241,009,345	&	$\dagger$	&	 1.061 	 & 	$\dagger$	\\	\\
A8	&	A71	&	32.01	&	$\dagger$	&	$\dagger$	&	391,461,947	&	$\dagger$	&	$\dagger$	&	$\dagger$	 & 	$\dagger$	\\	
&	A72	&	31.09	&	12,008.82	&	7,201.61	&	716,857,110	&	707,549,277	&	$\dagger$	&	 1.013 	 & 	$\dagger$	\\	
&	A73	&	34.43	&	11,991.83	&	7,201.44	&	810,760,974	&	797,302,300	&	$\dagger$	&	 1.017 	 & 	$\dagger$	\\	
&	A74	&	32.31	&	22,433.22	&	7,202.14	&	474,663,944	&	466,662,957	&	$\dagger$	&	 1.017 	 & 	$\dagger$	\\	
&	A75	&	34.41	&	19,625.22	&	7,205.37	&	617,667,994	&	604,037,347	&	$\dagger$	&	 1.023 	 & 	$\dagger$	\\	\hline
				\multicolumn{8}{l}{$^1$Mean of ten IHH runs with different random number seeds}\\
				\multicolumn{8}{l}{$^2$Ratio of mean IHH minimum cost to LPR and MIP solution values}\\
				\multicolumn{8}{l}{$\dagger$ No feasible solution found by CPLEX within two-hour time limit}\end{tabular} 
		} \end{center} \end{table}

Table \ref{probtimesH} provides computational results for the hub-network Test Set H.   Again, IHHO($\mathcal{P}$) found feasible
solutions to all problems and CPLEX did not find a feasible MIP solution for ten problems within the two-hour time limit. IHH costs averaged 2.7\% higher (median 1.9\%) than the non-integer LPR solutions and a mean of 2.3\% (median 1.4\%) above MIP solution values. But these high-quality IHH solutions were quickly identified.

\begin{table}[tbp]
	\begin{center}
		\caption{Test Set H Problems' Solution Times and Costs\label{probtimesH}}
		{\footnotesize
			\begin{tabular} {lcrrrrrrrr}
				
				&		&	 \multicolumn{3}{c}{Solution times, in seconds} 					&	 \multicolumn{3}{c}{Solution cost} 	& \multicolumn{2}{c}{IHH ratio$^2$ to:}				\\
				Grp	&	Prob	&	 IHH$^1$ 	&	 LPR 	&	 MIP 	&	 IHH$^1$ 	&	 LPR 	&	 MIP 	&LPR &MIP\\ \hline
H1	&	H1	&	0	&	0.18	&	29.42	&	29,955,889	&	28,539,832	&	28,618,951	&	 1.050 	 & 	 1.047 	\\	
&	H2	&	0	&	0.12	&	2.42	&	18,732,629	&	17,964,677	&	18,024,340	&	 1.043 	 & 	 1.039 	\\	
&	H3	&	0	&	0.04	&	0.57	&	22,622,073	&	22,422,453	&	22,447,019	&	 1.009 	 & 	 1.008 	\\	
&	H4	&	0	&	0.05	&	4.3	&	22,378,956	&	21,344,286	&	21,378,337	&	 1.048 	 & 	 1.047 	\\	
&	H5	&	0	&	0.11	&	2.32	&	6,558,798	&	5,935,733	&	5,960,416	&	 1.105 	 & 	 1.100 	\\	\\
H2	&	H11	&	0.01	&	0.19	&	5.58	&	70,538,736	&	68,777,857	&	68,980,048	&	 1.026 	 & 	 1.023 	\\	
&	H12	&	0.01	&	0.14	&	2.15	&	44,655,527	&	44,492,326	&	44,504,599	&	 1.004 	 & 	 1.003 	\\	
&	H13	&	0.01	&	0.15	&	0.68	&	54,413,060	&	54,189,697	&	54,260,908	&	 1.004 	 & 	 1.003 	\\	
&	H14	&	0.01	&	0.15	&	0.82	&	50,389,822	&	50,067,234	&	50,157,382	&	 1.006 	 & 	 1.005 	\\	
&	H15	&	0.01	&	0.11	&	2.41	&	13,944,509	&	13,726,647	&	13,764,490	&	 1.016 	 & 	 1.013 	\\	\\
H3	&	H21	&	0.36	&	9.43	&	159.31	&	144,381,818	&	142,649,463	&	143,598,565	&	 1.012 	 & 	 1.005 	\\	
&	H22	&	0.3	&	9.85	&	145.67	&	65,016,267	&	64,098,126	&	64,593,555	&	 1.014 	 & 	 1.007 	\\	
&	H23	&	0.21	&	3.54	&	57.74	&	80,029,905	&	79,211,507	&	79,348,400	&	 1.010 	 & 	 1.009 	\\	
&	H24	&	0.32	&	8.65	&	632.74	&	102,986,334	&	102,197,640	&	102,316,526	&	 1.008 	 & 	 1.007 	\\	
&	H25	&	0.39	&	9.79	&	148.4	&	104,165,646	&	102,197,678	&	102,722,378	&	 1.019 	 & 	 1.014 	\\	\\
H4	&	H31	&	0.85	&	30.96	&	922.14	&	327,687,154	&	325,837,074	&	326,010,526	&	 1.006 	 & 	 1.005 	\\	
&	H32	&	0.86	&	51.32	&	310.74	&	133,735,869	&	132,775,058	&	133,324,942	&	 1.007 	 & 	 1.003 	\\	
&	H33	&	0.79	&	12.2	&	47.12	&	191,168,482	&	190,474,085	&	190,602,451	&	 1.004 	 & 	 1.003 	\\	
&	H34	&	0.65	&	17.5	&	62.38	&	238,725,874	&	237,934,570	&	238,072,969	&	 1.003 	 & 	 1.003 	\\	
&	H35	&	0.57	&	21.39	&	72.11	&	233,146,966	&	232,279,190	&	232,412,534	&	 1.004 	 & 	 1.003 	\\	\\
H5	&	H41	&	0.05	&	1.87	&	15.51	&	69,673,389	&	65,287,854	&	65,742,141	&	 1.067 	 & 	 1.060 	\\	
&	H42	&	0.04	&	1.8	&	23.81	&	76,545,065	&	73,030,726	&	73,457,452	&	 1.048 	 & 	 1.042 	\\	
&	H43	&	0.05	&	1.61	&	37.88	&	60,697,045	&	57,730,345	&	58,155,068	&	 1.051 	 & 	 1.044 	\\	
&	H44	&	0.05	&	2.29	&	36.22	&	80,271,608	&	75,536,053	&	75,863,431	&	 1.063 	 & 	 1.058 	\\	
&	H45	&	0.04	&	2.53	&	39.21	&	62,502,443	&	59,563,649	&	59,901,249	&	 1.049 	 & 	 1.043 	\\	\\
H6	&	H51	&	0.17	&	10.55	&	183.8	&	156,275,156	&	150,945,216	&	151,690,295	&	 1.035 	 & 	 1.030 	\\	
&	H52	&	0.16	&	6.06	&	48.15	&	172,882,442	&	169,114,550	&	170,014,917	&	 1.022 	 & 	 1.017 	\\	
&	H53	&	0.2	&	6.13	&	72.72	&	141,093,921	&	137,522,791	&	138,217,544	&	 1.026 	 & 	 1.021 	\\	
&	H54	&	0.17	&	6.15	&	41.71	&	190,909,105	&	187,478,359	&	188,135,845	&	 1.018 	 & 	 1.015 	\\	
&	H55	&	0.16	&	9.08	&	44.59	&	153,301,857	&	149,465,335	&	150,208,616	&	 1.026 	 & 	 1.021 	\\	\\
H7	&	H61	&	8.95	&	1,202.56	&	7,201.74	&	161,927,789	&	153,606,071	&	$\dagger$	&	 1.054 	 & 	$\dagger$	\\	
&	H62	&	7.66	&	978.74	&	7,489.69	&	315,966,130	&	308,178,610	&	$\dagger$	&	 1.025 	 & 	$\dagger$	\\	
&	H63	&	7.94	&	1,249.44	&	7,201.77	&	355,177,320	&	343,644,420	&	$\dagger$	&	 1.034 	 & 	$\dagger$	\\	
&	H64	&	9.42	&	1,766.00	&	7,200.75	&	208,215,692	&	198,542,839	&	$\dagger$	&	 1.049 	 & 	$\dagger$	\\	
&	H65	&	8.24	&	2,285.61	&	7,200.69	&	275,247,104	&	264,016,690	&	$\dagger$	&	 1.043 	 & 	$\dagger$	\\	\\
H8	&	H71	&	26.36	&	9,806.65	&	7,202.07	&	391,242,593	&	383,969,884	&	$\dagger$	&	 1.019 	 & 	$\dagger$	\\	
&	H72	&	23.22	&	10,432.18	&	7,202.02	&	794,124,185	&	783,184,975	&	$\dagger$	&	 1.014 	 & 	$\dagger$	\\	
&	H73	&	24.48	&	13,263.41	&	7,202.30	&	881,780,074	&	868,298,779	&	$\dagger$	&	 1.016 	 & 	$\dagger$	\\	
&	H74	&	28.2	&	19,977.42	&	7,202.61	&	521,002,962	&	511,999,669	&	$\dagger$	&	 1.018 	 & 	$\dagger$	\\	
&	H75	&	24.76	&	18,301.86	&	7,202.07	&	691,022,207	&	678,169,628	&	$\dagger$	&	 1.019 	 & 	$\dagger$	\\	\hline
				\multicolumn{8}{l}{$^1$Mean of ten IHH runs with different random number seeds}\\
				\multicolumn{8}{l}{$^2$Ratio of mean IHH minimum cost to LPR and MIP solution values}\\
				\multicolumn{8}{l}{$\dagger$ No feasible solution found by CPLEX within two-hour time limit}\end{tabular} 
		} \end{center} \end{table}

The solution times for Test Sets A and H are  summarized  in Table \ref{agt}, where the 
IHH code's best, mean, and worst running times by problem
group are shown in columns 3--5.  The average ratio of
IHHO($\mathcal{P}$)'s running time to LPR and MIP running times are
shown in the last two columns (where feasible LPR or MIP solutions exist). 
For set A, the ratios indicate that the average IHH solution time is 172 times faster than the CPLEX LPR solver and 565 time faster than the CPLEX MIP code, which could not find an
integer solution for 20\% of the problems. 
The longest  IHH solve time for any combination of problem
and seed is 52.58 seconds for a problem with 17,772,480 binary
decision variables, 12,342 commodities, and 1,440 arcs.

The hub Test Set H gave similar results, with IHH solving these problems 129 times faster than LPR and 506 time quicker than the MIP solvers. Although the problem dimensions are the same as Test Set A, these seem to be easier problems for IHH, hence smaller solve times.

\begin{table}[tbp]
	\begin{center}
		\caption{Problem Group Solution Time: IHH Mean,  LPR and MIP Ratios\label{agt}}
		
		\begin{tabular} {lrrrc|rr}\\
			&\multicolumn{4}{c|}{IHH Running Time (sec)}&LPR:IHH & MIP:IHH \\
			\multicolumn{1}{c}{Group} & \multicolumn{1}{c}{Mean} & \multicolumn{1}{c}{Best} & \multicolumn{1}{c}{Worst} & Rank$^*$ 
			& \multicolumn{1}{|c}{time ratio$^1$}& \multicolumn{1}{c}{time Ratio$^2$ } \\
			\hline
			A1	&	0.02 	&0.00	&	 	1.00	&A&	 3.4 	&	 85.3 	\\
			A2	&	0.01 	&0.00	&	 	0.03	&A&	 466.9 	&	 150.5 	\\
			A3	&	0.74 	&0.46	&	 	1.57	&A,B&	 20.4 	&	 1,377.4 	\\
			A4	&	1.89 	&1.25	&	 	3.08	&B&	 60.2 	&	 1,171.3 	\\
			A5	&	0.06 	&0.04	&	 	0.11	&A&	 19.0 	&	 360.4 	\\
			A6	&	0.18 	&0.11	&	 	0.37	&A&	 37.7 	&	 248.8 	\\
			A7	&	11.06 	&6.66	&	 	20.76	&C&	 267.5 	&	 $\dagger$ 	\\
			A8	&	32.85 	&22.24	&	 	52.58	&D&	 502.7 	&	 $\dagger$ 	\\ \hline
			\multicolumn{2}{l}{Average}       &           &           &&    172.2     &   565.6 \\ \\ 
			H1	&	0.00 	& 0.00 	&	 	 0.01 	&A&	 $\infty$ 	&	 $\infty$ 	\\
			H2	&	0.01 	& 0.00 	&	 	 0.03 	&A&	  14.8 &	 232.8\\
			H3	&	0.32 	& 0.17 	&	 	 0.77 	&A& 25.8 &	 715.0	\\
			H4	&	0.74 	& 0.48 	&	 	 1.97 	&A&	  36.0 &	 382.3	\\
			H5	&	0.05 	& 0.03 	&	 	 0.07 	&A&	  40.4 &	 610.5\\
			H6	&	0.17 	& 0.12 	&	 	 0.26 	&A&	  44.7 &	 460.0\\
			H7	&	8.44  	& 6.37 	&	 	 13.61	&B&	  177.3&	 860.1 	\\
			H8	&	25.40 	& 18.77	&	 	 38.58	&C&	  565.2&	 283.6 	\\ \hline
			\multicolumn{2}{l}{Average}       &           &           &&    129.2     &   506.3 \\  
			
			\multicolumn{6}{l}{\scriptsize{ * - Tukey's Significant Difference test ranking}} \\
			\multicolumn{6}{l}{\scriptsize{ 1 - (Mean LPR Solution Time)/(Mean IHH Solution Time)}} \\
			\multicolumn{6}{l}{\scriptsize{ 2 - (Mean MIP Solution Time)/(Mean IHH Solution Time)}} \\
			\multicolumn{6}{l}{\scriptsize{ $\dagger$ - No feasible solution found by CPLEX12 within 7200 seconds.}} \\
		\end{tabular}  \end{center} \end{table}

\subsection{Statistical Analyses}

Statistical analysis of results is performed using SAS Version 9.4 to test whether IHHO($\mathcal{P}$)'s performance
with respect to time and solution quality is significantly
different for the various problem groups. As solutions are not available
from LPR and MIP for all problems, the least-squares GLM Procedure is used to analyze such unbalanced data.  The
analysis reveals whether the differences between the observed
means of populations, the groups, are statistically significant.
Using Tukey's Significan Difference Test, each population is given a letter representing its ranking.
Populations with the same letter do not have statistically
significant differences between their means. More than one letter
indicates a population's mean is not significantly different than
the means of more than one distinct set of populations. Members
labelled ``A'' have the best values, lower objective function or
running times. Values become progressively worse in alphabetical
order.

Table \ref{agt}'s column 5 gives the ranking of problem groups within each test set based on solution time. The most difficult set A problem groups were A7 and A8, denoted by their C and D rankings. Similarly, for set H, groups H7 and H8 had the longest solution times, however the times for groups H1--H6 were not significantly different.

Table \ref{agrc} shows the ratio of the IHHO($\mathcal{P}$) and
 LPR and MIP objective function values.  These ratios indicate the percentage difference between the IHH
 solution value and the corresponding value of the linear programming relaxation solution or the MIP 
results. For example, a ratio of 1.036 shows that the IHH solution cost averaged 3.6%
larger than CPLEX12's linear programming relaxation or its mixed integer programming solution value.
For Test Set A, IHHO($\mathcal{P}$)'s costs averaged 3.5\% higher than the LPR solution value but, as noted above, this was found 172 times faster. Similarly, the IHH costs averaged 3.1\% higher than the MIP values, but were found 565 times quicker, per Table \ref{agt}.
The Tukey rankings in column 5 do not reveal an obvious pattern as to what might make some problems more difficult.

To further explore the problem characteristics that affect IHHO($\mathcal{P}$) solution times, a regression analysis was performed based on data from Test Sets A and H. The observed solution times was the dependent variable and the explanatory variables from Table \ref{agc}: $|N|, |A|, |K|$, number of binary variables and constraints, average cost, average demand, average arc capacity, and the network topology (hub or non-hub).  The regression has an $r^2=0.9795$. Only three of the nine explanatory variables are not statistically significant: capacity, cost, and degree. Of the six significant predictors, solution time increased with $|A|$ and the number of problem constraints, but decreased if a hub topology was used, or if $|N|, |K|$, or the number of binary variables increased. 

\begin{table}[tbp]
	\begin{center}
		\caption{IHH Solution Value Ratio to LPR, MIP\label{agrc}}
		\begin{tabular} {lrrrc|rrrc} \\
			&\multicolumn{4}{c|}{IHH:LPR Cost Ratio$^1$}&\multicolumn{3}{c}{IHH:MIP Cost Ratio$^2$} \\
			\multicolumn{1}{c}{Group} & \multicolumn{1}{c}{Mean} & 
			\multicolumn{1}{c}{Best} & \multicolumn{1}{c}{Worst} & 
			\multicolumn{1}{c|}{Rank}& 
			\multicolumn{1}{c}{Mean} & \multicolumn{1}{c}{Best} & \multicolumn{1}{c}{Worst}  \\
			\hline
			A1	& 1.036 & 1.027 & 1.048 & A,B & 1.029 & 1.021 & 1.041   \\
			A2	& 1.017 & 1.012 & 1.024 & A,B & 1.015 & 1.010 & 1.021   \\
			A3	& 1.060 & 1.055 & 1.067 & B & 1.052 & 1.047 & 1.059   \\
			A4	& 1.021 & 1.020 & 1.023 & A,B & $\dagger$ & $\dagger$ & $\dagger$    \\
			A5	& 1.064 & 1.047 & 1.080 & B & 1.056 & 1.040 & 1.072   \\
			A6	& 1.017 & 1.013 & 1.020 & A & 1.013 & 1.010 & 1.017   \\
			A7	& 1.047 & 1.046 & 1.050 & A,B & $\dagger$	 & 	 $\dagger$	 & 	  $\dagger$  \\
			A8	& 1.017 & 1.017 & 1.018 & A,B & $\dagger$ 	&	$\dagger$ 	&$\dagger$ 	   \\ \hline 
			Average & 1.035 & 1.030 & 1.042 &	&1.031 &  1.024 & 1.038     \\
			\\
			H1	& 1.051 & 1.039 & 1.064 & C,B & 1.048 & 1.036 & 1.061    \\
			H2	& 1.011 & 1.009 & 1.014 & A & 1.009 & 1.007 & 1.012    \\
			H3	& 1.013 & 1.011 & 1.015 & A & 1.008 & 1.006 & 1.010    \\
			H4	& 1.005 & 1.004 & 1.005 & A & 1.003 & 1.003 & 1.004    \\
			H5	& 1.056 & 1.044 & 1.066 & C & 1.049 & 1.037 & 1.060    \\
			H6	& 1.026 & 1.019 & 1.034 & B,A & 1.021 & 1.014 & 1.029   \\
			H7	& 1.041 & 1.039 & 1.044 &  &  $\dagger$ 	& $\dagger$ 	& $\dagger$ 		 \\
			H8	& 1.017 & 1.016 & 1.018 &  &  $\dagger$ 	&$\dagger$ 	&$\dagger$ 		 \\  \hline
			Average & 1.027 & 1.022 & 1.033 &	&1.023 &	 1.017 	&1.029 	     \\
			\multicolumn{8}{l}{\scriptsize{ 1 - (IHH Objective Function Value)/(LPR Objective Function Value)}} \\
			\multicolumn{8}{l}{\scriptsize{ 2 - (IHH Objective Function Value)/(MIP Objective Function Value)}} \\
			\multicolumn{8}{l}{\scriptsize{ $\dagger$ - No feasible solution found by CPLEX12 within 7200 seconds.}} \\
		\end{tabular}   \end{center} \end{table}

Additional analysis was performed on results of test set A to determine the impact of the $hm(\lambda_k)$ function on overall performance. As noted in Section \ref{routingdecision} and \ref{asymptotic}, $hm(\lambda_k)$ is designed to force IHHO($\mathcal{P}$) to converge to an equilibrium if the market forces are insufficient. The code was modified to count the number of times the hurdle multiplier prevented a commodity from switching to a new route so when $rc(CP(k),k,A,\mathbf X)\geq hm(\lambda_k)rc(CP(k),k,A,\mathbf X)$ while $rc(NewPath,k,A,\mathbf X) < rc(CP(k),k,A,\mathbf X)$. For the 400 runs of IHHO($\mathcal{P}$) performed the hurdle multiplier had an effect during five calls to Route($k, \lambda_k, \mathcal{P}$). These five instances occurred for one seed of one problem in Group A8. These results indicate that for most problems and commodities IHH's price mechanism alone is sufficient to reach an equilibrium.

Analysis was also performed to determine the impact of the FeasPath($\mathcal{P}$) procedure on overall performance. As noted in Section \ref{feaspath}, the marginal market cost curve is not an exact match to the original ODIMCF problem and FeasPath($\mathcal{P}$) attempts to close the gap by finding lower original cost paths to fully utilize arc capacities and by rerouting commodities that paid a large marginal market cost to exceed an arc's capacity constraint. For all runs of problem set A, 69.25\% of solutions after IHHO($\mathcal{P}$)'s main loop are feasible with respect to arc capacity constraints. For the infeasible solutions, the mean and median percentage of arc capacity constraints violated are 0.75\% and 0.56\% respectively with FeasPath($\mathcal{P}$) able to resolve all violations. For the feasible solutions, the mean and median percentage of ODIMCF objective function improvement after FeasPath($\mathcal{P}$) are 0.66\% and 0.70\% respectively. For all runs, the mean and median number of reroutes per commodity during FeasPath(P) are 0.090 and 0.089 respectively indicating most commodities do not change routes during FeasPath($\mathcal{P}$). These results indicate that the main loop of the IHHO($\mathcal{P}$) algorithm based on the invisible hand analogy is doing most of the work with FeasPath($\mathcal{P}$) closing the small remaining gap.

\subsection{Large Problems: Test  Set L}
Test Set L contained the largest ODIMCF problems solved to explore the capabilities of the IHH algorithm. Since these problems are beyond the solvability of CPLEX, they are only run using IHHO($\mathcal{P}$). The problems dimensions are given in Table \ref{agc} with  Table \ref{setLtimes} providing solution times, costs, and coefficients of variation from the computer experiments. These contain the largest ODIMCF problem instances published to date, with group L6 networks containing 1,920 nodes, 23,040 arcs, and 106,338 commodities, as can be found in industrial problems \cite{yasukawa2009rfc}. 

The results show that the IHH can even solve problems with over two billion binary variables and 200 million constraints in 3,200 seconds. If the results from Test Sets A and H continue to hold, the solution values could be within a few percentage points of an exact solution's. 

\begin{table}[tbp]
	\begin{center}
		\caption{Large Problem Set L: Solution Time, Cost, Coefficient of Variation \label{setLtimes}} 
		\begin{tabular} {llrrr} \\
			Group	&	Problem	&	IHH time	&	IHH cost	&	CV cost	\\ \hline
			L1	&	L51 	&	 77.78 	&	 1,426,590,347 	&	 0.00018 	 \\
			&	L52 	&	 81.73 	&	 1,254,903,691 	&	 0.00013 	 \\
			&	L53 	&	 73.63 	&	 1,703,192,064 	&	 0.00014 	 \\
			&	L54 	&	 81.48 	&	 1,955,056,175 	&	 0.00017 	 \\
			&	L55 	&	 72.03 	&	 1,434,497,785 	&	 0.00013 	 \\ \cline{2-5}
			&	Average:	&	 77.33 	&		&	 0.00015 	 \\
			&		&		&		&		 \\
			L2	&	L56 	&	 226.81 	&	 3,562,506,827 	&	 0.00002 	 \\
			&	L57 	&	 213.14 	&	 3,123,060,684 	&	 0.00005 	 \\
			&	L58 	&	 218.98 	&	 4,236,881,946 	&	 0.00004 	 \\
			&	L59 	&	 205.03 	&	 4,861,652,396 	&	 0.00004 	 \\
			&	L60 	&	 208.00 	&	 3,552,577,022 	&	 0.00004 	 \\ \cline{2-5}
			&	Average:	&	 214.39 	&		&	 0.00004 	 \\
			&		&		&		&		 \\
			L3	&	L71 	&	 286.58 	&	 3,660,398,107 	&	 0.00012 	 \\
			&	L72 	&	 294.18 	&	 3,793,204,910 	&	 0.00011 	 \\
			&	L73 	&	 287.23 	&	 2,713,212,910 	&	 0.00013 	 \\
			&	L74 	&	 275.33 	&	 4,116,477,552 	&	 0.00008 	 \\
			&	L75 	&	 284.63 	&	 3,766,807,718 	&	 0.00010 	 \\ \cline{2-5}
			&	Average:	&	 285.59 	&		&	 0.00011 	 \\
			&		&		&		&		 \\
			L4	&	L76 	&	 820.17 	&	 9,119,420,821 	&	 0.00003 	 \\
			&	L77 	&	 764.44 	&	 9,397,310,591 	&	 0.00002 	 \\
			&	L78 	&	 813.81 	&	 6,720,114,699 	&	 0.00002 	 \\
			&	L79 	&	 820.91 	&	 10,235,141,998 	&	 0.00004 	 \\
			&	L80 	&	 783.75 	&	 9,318,572,755 	&	 0.00003 	 \\ \cline{2-5}
			&	Average:	&	 800.62 	&		&	 0.00003 	 \\
			&		&		&		&		 \\
			L5	&	L91 	&	 1,114.24 	&	 6,497,960,761 	&	 0.00008 	 \\
			&	L92 	&	 1,222.81 	&	 20,287,326,224 	&	 0.00007 	 \\
			&	L93 	&	 1,132.24 	&	 8,091,918,047 	&	 0.00007 	 \\
			&	L94 	&	 1,141.31 	&	 15,842,224,803 	&	 0.00009 	 \\
			&	L95 	&	 1,156.20 	&	 8,657,046,549 	&	 0.00006 	 \\ \cline{2-5}
			&	Average:	&	 1,153.36 	&		&	 0.00007 	 \\
			&		&		&		&		 \\
			L6	&	L96 	&	 3,118.31 	&	 16,154,800,326 	&	 0.00003 	 \\
			&	L97 	&	 3,210.38 	&	 50,496,932,124 	&	 0.00003 	 \\
			&	L98 	&	 3,306.78 	&	 20,132,144,074 	&	 0.00002 	 \\
			&	L99 	&	 3,141.18 	&	 39,421,362,436 	&	 0.00002 	 \\
			&	L100 	&	 3,181.98 	&	 21,458,101,592 	&	 0.00003 	 \\ \cline{2-5}
			&	Average:	&	 3,191.72 	&		&	 0.00002 	 \\ 
		\end{tabular}  \label{largestat} \end{center} \end{table}

To assess the impact of the inherent randomness in IHH, each problem was solved ten times with different random number seed values and the average time and cost reported. An evaluation of the computational results of multiple runs per problem shows that the mean coefficient of variation (standard deviation normalized by the mean) for test sets A and H of IHH solution values is 0.35\% (median is 0.00201), reflecting small variation in the resulting solution values; for the largest set L, the even smaller average CV = 0.00007, shown in Table \ref{setLtimes}. This lack of variation indicates that the IHH provides robust results for these problems.

\section{Conclusions}

Origin-destination integer multicommodity flow problems occur in a variety of application areas---including logistics and telecommunications---and are often of large
dimensions, in terms of nodes, arcs,  commodities, binary variables, and constraints.  This invisible-hand heuristic, inspired by the efficient economic processes underlying a market economy, provides a new method for solving large-scale instances of
such problems. Computational testing demonstrates it has the capability
to achieve integer feasible solutions with excellent solution
quality relative to objective function value.

The current implementation's  time performance appears competitive over a wide range of problem characteristics.
Linear space requirements combined with fast running times enables
IHH to solve realistic problems with millions of constraints and hundreds of millions of binary variables that are beyond the reach of
other methods. The methodology is shown to identify high-quality integer solutions quickly, as verified in comparisons with state-of-the-art commercial software.

Further research into this approach could take advantage of parallel computing implementations whereby a host of competing system processes could emulate the distributed decision-making of a marketplace. By replacing randomized choices with algorithmic race conditions, the method might converge faster while still uncovering high-quality solutions and enable the solution of even larger problem instances. Additional research directions could evaluate application of this approach to variations of the ODIMCF problem. One variation to examine is problems where the underlying network does not have sufficient capacity to accommodate all commodities and a decision on which commodities to service must be made. Such research could include reporting on the state of the problem to inform decisions on future changes to the network. A second variation to examine is problems where there is a limit on the total route length for commodities and, a  third, is one with a demand schedule for commodities so that demand varies as a function of time possibly with commodities completely deactivating during certain times.

\bibliographystyle{acm}
\bibliography{ihh}
\end{document}